\documentclass[lettersize,journal]{IEEEtran}
\usepackage{amsmath,amsfonts}

\usepackage{array}
\usepackage{textcomp}
\usepackage{stfloats}
\usepackage{url}
\usepackage{verbatim}
\usepackage{graphicx}
\usepackage{cite}
\usepackage{amsmath,amsfonts}

\usepackage{array}
\usepackage{textcomp}
\usepackage{stfloats}
\usepackage{url}
\usepackage{verbatim}

\usepackage{subfigure}
\usepackage{multirow}

\usepackage{algorithm,algpseudocode}
\usepackage{algpseudocode}

\newtheorem{proposition}{Proposition}
\usepackage{caption}

\usepackage{braket}

\usepackage{booktabs}
\usepackage{amssymb}
\usepackage{graphicx}

\begin{document}

\title{Sharpness-Aware Cross-Domain Recommendation \\ to Cold-Start Users}

\author{Guohang Zeng, Qian Zhang,~\IEEEmembership{Member,~IEEE},  Guangquan Zhang, Jie Lu*, \IEEEmembership{Fellow,~IEEE}

        % <-this % stops a space
%\thanks{The work presented in this paper was supported by the Australian Research Council (ARC) under Laureate project FL190100149 and discovery project DP220102635.}

\thanks{This work has been submitted to the IEEE for possible publication. Copyright may be transferred without notice, after which this version may no longer be accessible.}

\thanks{The authors are with the Decision Systems and e-Service Intelligence Laboratory, Australian Artificial Intelligence Institute (AAII), Faculty of Engineering and Information Technology, University of Technology Sydney, Ultimo, NSW 2007, Australia (e-mail: guohang.zeng@student.uts.edu.au;  qian.zhang@uts.edu.au; guangquan.zhang@uts.edu.au; jie.lu@uts.edu.au)}% <-this % stops a space
\thanks{* Corresponding author}}

%\markboth{IEEE TRANSACTIONS ON SYSTEMS, MAN, AND CYBERNETICS: SYSTEMS}%
\markboth{}%
{Shell \MakeLowercase{\textit{et al.}}: A Sample Article Using IEEEtran.cls for IEEE Journals}



\maketitle

\begin{abstract}
Cross-Domain Recommendation (CDR) is a promising paradigm inspired by transfer learning to solve the \textit{cold-start problem} in recommender systems. Existing state-of-the-art CDR methods train an explicit mapping function to transfer the cold-start users from a data-rich source domain to a target domain. However, a limitation of these methods is that the mapping function is trained on overlapping users across domains, while only a small number of overlapping users are available for training. By visualizing the loss landscape of the existing CDR model, we find that training on a small number of overlapping users causes the model to converge to sharp minima, leading to poor generalization. Based on this observation, we leverage loss-geometry-based machine learning approach and propose a novel CDR method called Sharpness-Aware CDR (SCDR). Our proposed method simultaneously optimizes recommendation loss and loss sharpness, leading to better generalization with theoretical guarantees. Empirical studies on real-world datasets demonstrate that SCDR significantly outperforms the other CDR models for cold-start recommendation tasks, while concurrently enhancing the model's robustness to adversarial attacks.
\end{abstract}


\section{Introduction}

\IEEEPARstart{R}{ecommender} systems are predicated on collaborative filtering, which aims to learn latent representations for users and items through their historical interactions \cite{bobadilla2013recommender,koren2021advances, rafailidis2015modeling, wu2019deep}. In this latent representation space, users and items with similar preferences are close to each other, allowing us to recommend items to users based on their proximity. However, it is challenging to make personalized recommendations for new users who have no interaction history with items. This issue, known as the \textit{cold-start problem} \cite{lam2008addressing}, can result in a lack of personalized recommendations for new platform users due to insufficient data to derive their representations through collaborative filtering. In recent years, substantial efforts have been made to overcome the cold-start problem, and one of the most promising solutions is Cross-Domain Recommendation (CDR) \cite{lika2014facing,cremonesi2011cross, zhang2021deep}.

\begin{figure}[t]
\centerline{\includegraphics[width=0.45\textwidth]{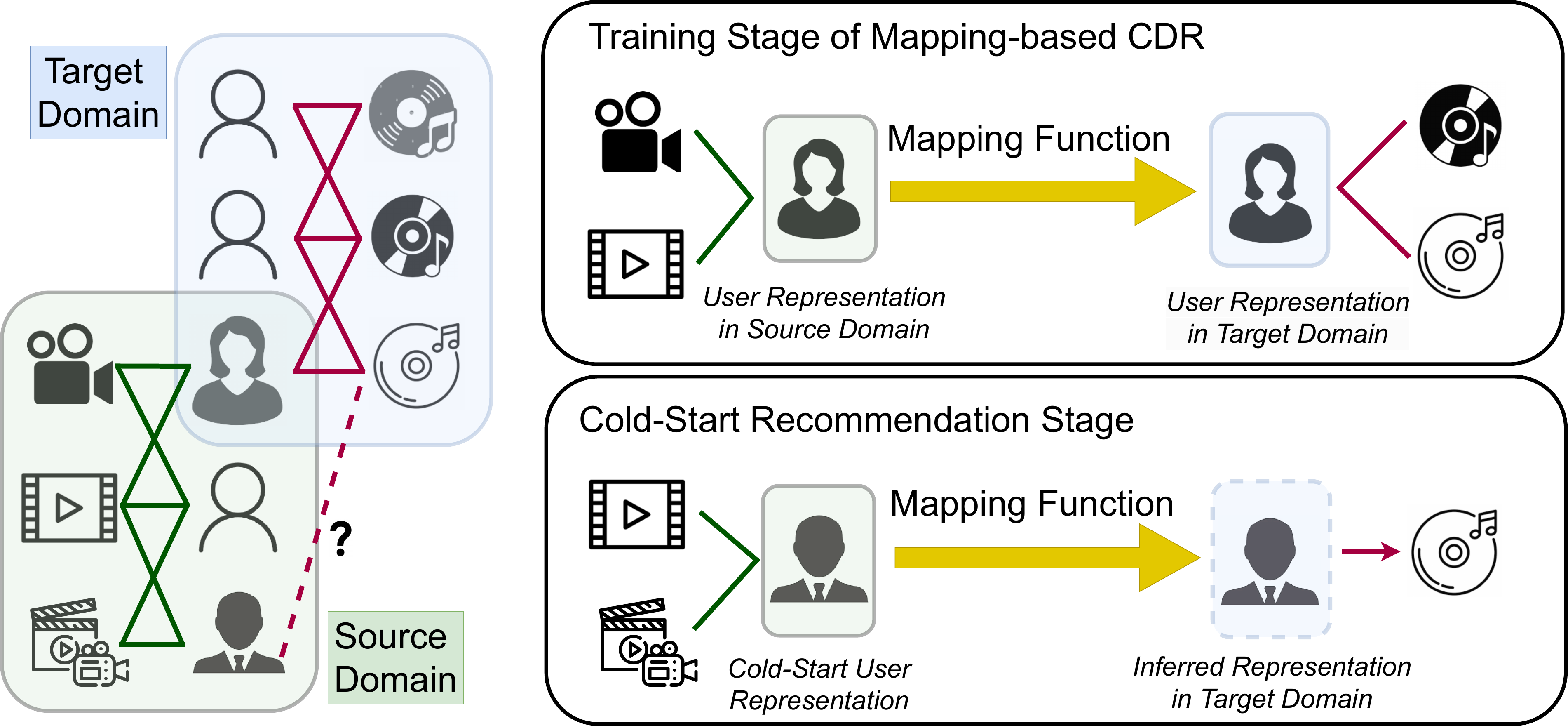}}
\caption{An example of the EMCDR-based method: it learns an explicit mapping function for users across domains to obtain the representation of cold-start users in the target domain.}
\label{fig-mm}
\end{figure}

CDR leverages the concept of transfer learning, utilizing data from a data-rich source domain to acquire latent representations for cold-start users in a target domain. Among various CDR methods, the Embedding and Mapping Cross-Domain Recommendation approach (EMCDR) \cite{man2017cross} has emerged as a successful paradigm for transferring knowledge across domains. EMCDR initially pre-trains the latent representations for each domain and subsequently learns an explicit \textit{mapping function} between source and target domains by minimizing the distance between the overlapping users across domains. As shown in Figure \ref{fig-mm}, the mapping function establishes a connection between source and target domains, enabling recommendations for cold-start users whose representation in the target domain can be inferred using the mapping function. In recent years, several EMCDR-based methods \cite{man2017cross,kang2019semi,zhu2022personalized} have achieved state-of-the-art performance on CDR cold-start recommendation tasks.

However, the EMCDR-based methods exhibit several limitations. Firstly, the mapping function relies heavily on overlapping users that have interaction histories in both the source and target domains, yet the presence of such overlapping users is often limited. Taking the Amazon dataset \cite{he2016ups} as an example, \cite{kang2019semi} found that the average ratio of overlapping users relative to the total user across any two domains is less than 5\%. Training EMCDR-based methods with such a small percentage of users can lead to biased results. Secondly, current EMCDR-based methods primarily focus on optimizing the parameters of the mapping function. However, given the heterogeneous latent representation in the source and target domains, as pointed out by \cite{liu2021leveraging}, optimizing the mapping function based on a limited number of overlapping users may not be sufficient for a CDR model to generalize well.

\begin{figure}[t]
\centering
    \subfigure[EMCDR]{
        \includegraphics[width=0.44\columnwidth]{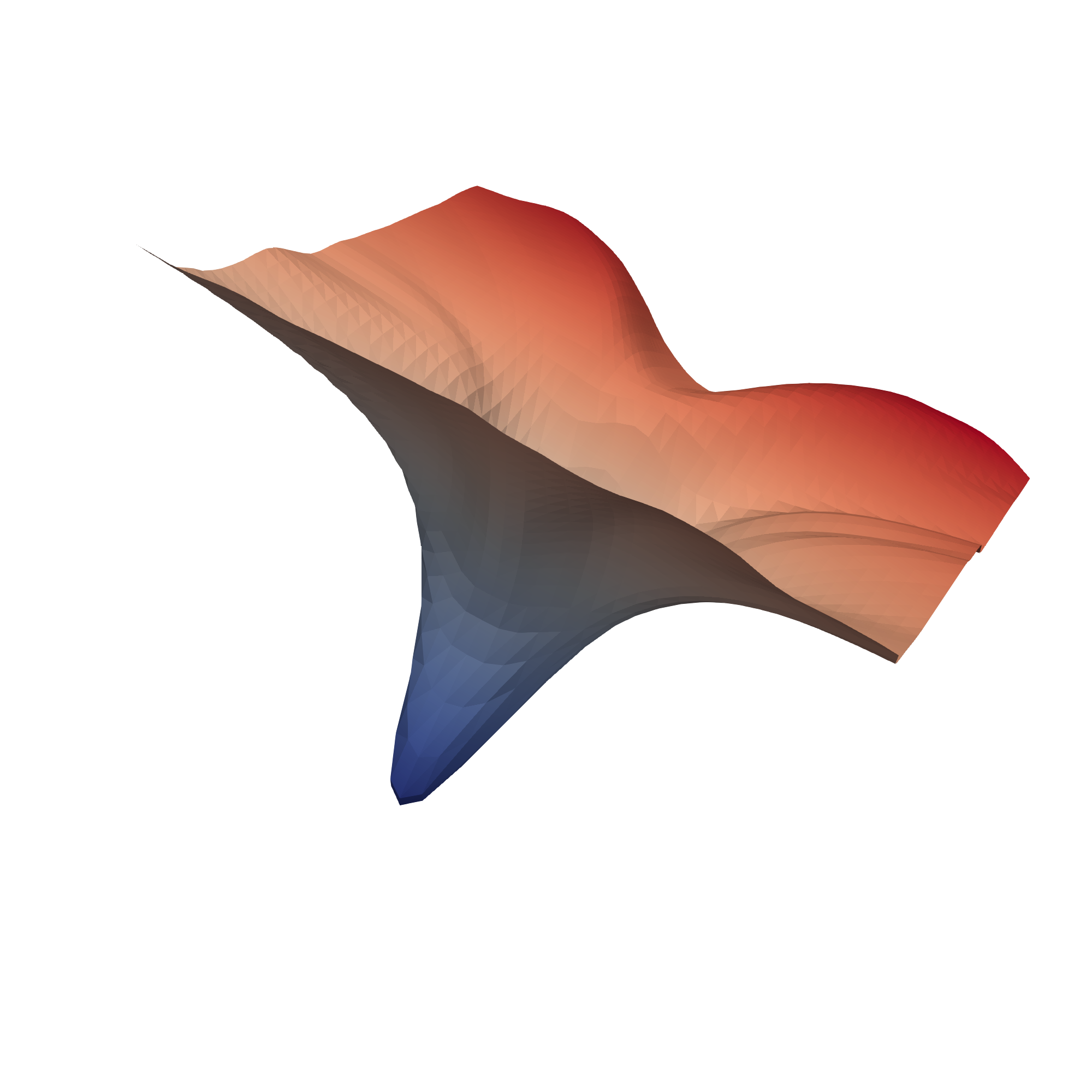}
    }
    \subfigure[SCDR]{
    \raisebox{-20pt}[0pt][0pt]
    
        \includegraphics[width=0.44\columnwidth]{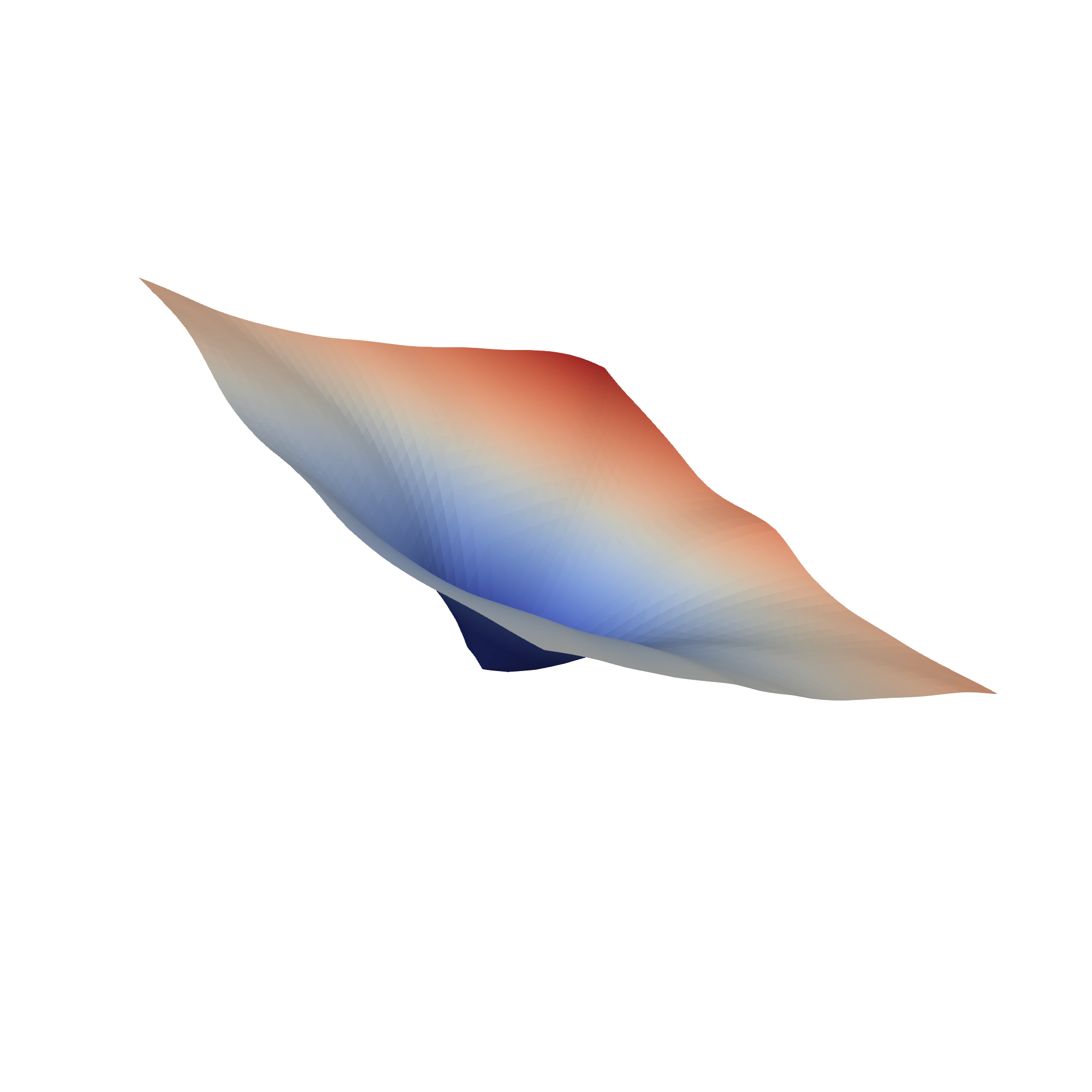} 
        \vspace{-10em}
    }
\vspace{-0.0 in}
\caption{Visualization of loss landscape for user representation space. Compared with the EMCDR\cite{man2017cross} with a sharp loss landscape, the proposed SCDR converges to a flatter loss landscape on Amazon CDR dataset (Movie $\rightarrow$ Music).}
\label{fig:landscape}
\end{figure}

To validate this issue, we investigated the loss landscape of the EMCDR model trained on the Amazon dataset. We adopted the visualization method proposed by \cite{li2018visualizing}, calculating loss values by moving the model parameters along two random directions to generate a 3D loss landscape from a high dimension representation space. By visualizing the loss landscapes of EMCDR in the user representation space, we find that EMCDR converges to sharp local minima as shown in Figure \ref{fig:landscape} (a). These minima are characterized by unstable curvatures, as the loss values change rapidly in their vicinity. As discussed by recent studies \cite{huang2023boosting}, such sharp local minima are associated with several undesirable properties, including poor generalization, and vulnerability to adversarial attacks \cite{wu2020adversarial}.

Inspired by the connection between loss landscape and generalization \cite{foret2020sharpness}, we propose the Sharpness-Aware Cross-Domain Recommendation (SCDR) method,  which optimizes both the loss function and the geometric properties of the model’s loss landscape simultaneously. SCDR optimizes the CDR model to converge to a flatter minima through a min-max optimization approach, meaning that neighbourhoods of overlapping user representations have uniformly low loss. Intuitively, SCDR encourages the maintenance of consistent preferences within the $\ell_2$-norm ball of the representation space for overlapping users, thus avoiding the bias associated with merely learning the mapping function from the scarce overlapping users. A notable difference between SCDR and previous mapping-based methods in CDR is that we do not just optimize the mapping function, but also aim to optimize a user latent representation that is more favourable for the CDR task. As shown in Figure \ref{fig:landscape} (b), the model trained with SCDR converges to a much flatter loss landscape. Experiments indicate that our method outperforms existing CDR methods and achieves state-of-the-art results. Additionally, we found that flattening the loss landscape makes the CDR model less susceptible to gradient-based adversarial attacks, thereby enhancing the model's robustness. 

In summary, our contributions are as follows:
\begin{itemize}
    \item We propose a novel method named Sharpness-Aware Cross-Domain Recommendation (SCDR) to address the cold-start problem. Given that the number of overlapping users is limited, which leads to the previous methods easily converging to sharp minima, the proposed SCDR method encourages the model to converge to flat minima with better generalization.
    \item We validated the effectiveness of our method on three cross-domain recommendation tasks from the Amazon dataset. Comprehensive experiments show that our method significantly outperforms other baseline methods and achieves state-of-the-art performance. An ablation study confirms the effectiveness of the proposed SCDR method.
    \item We verified that SCDR can effectively defend against white-box adversarial attacks, indicating that SCDR improves the model's robustness while converging to flat minima.
\end{itemize}

The rest of this article is organized as follows. In Section II, we review cross-domain recommendation methods and sharpness-aware minimization. The cross-domain cold-start recommendation problem and related preliminary settings are formulated in Section III. Motivated by the challenges in current cross-domain recommendation methods, our proposed SCDR method is introduced in detail in Section IV. In Section V, we evaluate our proposed method using real-world datasets, including 3 CDR scenarios from the Amazon dataset and the Douban dataset. We analyze many properties of the proposed method, including its ability to enhance adversarial robustness, and converge to flat minima. Finally, we conclude our work and discuss future studies in Section VI.

\section{Related Work}

In this section, we will review cold-start recommendation and cross-domain recommendation methods, then present a literature survey about sharpness-aware minimization with its motivation.

\subsection{Cold-Start Recommendation }

Cold-start recommendation is a critical challenge in recommender systems, particularly when dealing with new users or items that have little to no historical data available \cite{lika2014facing}. In such scenarios, traditional collaborative filtering methods struggle to provide accurate recommendations due to the lack of user- item interactions \cite{son2016dealing}. The cold-start problem can be divided into two categories: the new user cold- start problem and the new item cold-start problem. The new item cold-start problem occurs when a new item has just been registered on a system, and can be partially resolved through technical means, such as having staff members provide prior ratings for the new item. On the other hand, the new user cold-start problem is more challenging because no prior ratings can be made due to privacy considerations. Therefore, addressing the new user cold-start problem is a more critical issue in recommendation systems; in this paper, we focus solely on the new user cold-start problem.

To address this issue, various approaches have been proposed to tackle the new user cold-start problem in recommender systems. One common strategy is to leverage side information, such as user demographics, item attributes, or social network data, to enhance the recommendation process \cite{gope2017survey}. By incorporating additional contextual information, the system can infer preferences for new users based on their similarities with existing entities \cite{zhu2017chrs}. For example, \cite{almazro2010survey}  user demographics information can be utilized as an additional data source to provide movie recommendations to cold-start users. \cite{poirier2010reducing} leveraged blog textual data to construct a user-item rating matrix and establish recommendations for cold-start users. \cite{sun2011content} clusters users based on the user-item rating matrix and trains a decision tree using the clustering results and user demographics information to provide recommendations for new users. However, in the absence of side information, the aforementioned methods cannot be used to solve the cold-start problem. Under such circumstances, other paradigms for addressing the cold-start problem have been proposed, such as cross-domain recommendation.

\subsection{Cross-Domain Recommendation} 

Cross-domain recommendation \cite{lu2015recommender,zhu2021cross,wang2021cross} utilizes the concept of domain adaptation, aiming to harness knowledge from the source domain of recommendation to aid in training a collaborative filtering model for the target domain, in an attempt to overcome data scarcity issue and cold-start problems. A straightforward solution of CDR is to jointly learn the representations of both domains. \cite{singh2008relational} proposed Collective Matrix Factorization, which performs matrix factorization simultaneously on data from the two domains to transfer knowledge from the source domain to the target domain. Subsequent work revolves around designing better DNNs to improve the performance of CDR tasks. \cite{hu2018conet} designed a neural network structure that allows the interaction of information from both domains to overcome data sparsity issues. \cite{zhu2019dtcdr} proposed a dual-target CDR model to improve recommendation performance simultaneously in both richer and sparser domains. \cite{zhu2021graphical}) utilized a graphical and attention framework to further enhance the interaction of information between different domains. \cite{xie2022contrastive} employed contrastive learning approach to enhance Sequence Learning for Cross-Domain Recommendation. \cite{zhang2021deep} used an adversarial network to align latent feature spaces between the source and target domains.

Another paradigm of CDR approaches is mapping-based CDR. Compared to other CDR paradigms, mapping-based CDR methods pay more attention to solving the cross-domain cold-start problem. \cite{man2017cross} explicitly learn a mapping function between the source domain and the target domain, using a mapping function to map cold-start users to the target domain for initialization.  \cite{zhu2022personalized} used a meta-network to generate a personalized bridge function to achieve personalized transfer of preferences for cold-start users. \cite{chen2022clcdr} simultaneously utilized the information about overlapping users and user-item interactions through contrastive learning to enhance the performance of cross-domain cold-start recommendations. Given the scarcity of shared users, \cite{kang2019semi} employed semi-supervised methods to utilize non-shared data to learn the mapping function as a data augmentation strategy. In this work, we consider overcoming the issue of limited training data for overlapping users from the perspective of the properties of the loss landscape.

\subsection{Sharpness-Aware Minimization}

The relationship between the generalization gap and the geometric properties of the loss landscape has been studied for a long time in the machine learning community \cite{dziugaite2017computing, dinh2017sharp, keskar2016large, hochreiter1997flat}. \cite{jiang2019fantastic} discovered that sharpness-based measures have the highest correlation with generalization via extensive empirical studies, which inspired \cite{foret2020sharpness} to enhance generalization ability by directly penalizing loss sharpness. \cite{foret2020sharpness} proposed Sharpness-Aware Minimization (SAM), which aims to simultaneously minimize the loss value and loss sharpness and seeks parameters that reside in neighborhoods with uniformly low loss. Recent studies have expanded and explored the properties of SAM from many perspectives, including generalization bounds \cite{wen2022sharpness}, computational efficiency \cite{liu2022towards} \cite{du2022sharpness}, and its relationship with adversarial robustness \cite{wei2023sharpness}. Recently, SAM has also been introduced into recommendation systems to enhance the performance of sequence recommendation \cite{lai2023enhancing} and graph recommendations \cite{chen2023sharpness}. In this work, we are the first to introduce SAM into CDR to address the generalization issues associated with cold-start recommendations.

\section{Preliminaries}

The CDR scenario for cold-start user recommendation involves two distinct domains: namely, the source domain and the target domain. Each domain comprises three parts: the user set $\mathcal{U}$, the item set $\mathcal{V}$, and the interaction matrix $\mathbf{R}$. Specifically, the user set $\mathcal{U}$ can be represented as $\left\{u_1, u_2, \ldots\right\}$, and the item set $\mathcal{V}$ is $\left\{v_1, v_2, \ldots\right\}$, where $u_i$ and $v_j$ represent the $i$-th user and $j$-th item, respectively. It is important to note that $u$ and $v$ are categorical variables used to denote their respective identifiers. The interaction matrix $\mathbf{R}$ is an $\left|\mathcal{U}\right| \times \left|\mathcal{V}\right|$ matrix, which can be represented as $[R_{ij}]$, where $R_{ij} \in \mathbb{Z}$ is the interaction between $u_i$ and $v_j$. We denote the source domain and target domain as $\left\{\mathcal{U}^{s}, \mathcal{V}^{s}, \mathbf{R}^{s}\right\}$ and $\left\{\mathcal{U}^{t}, \mathcal{V}^{t}, \mathbf{R}^{t}\right\}$, respectively. We also assume that there are overlapping users across the domains, denoted as $\mathcal{U}^o=\mathcal{U}^s \cap \mathcal{U}^t$, while there are no shared items between the domains, i.e., $\mathcal{V}^s \cap \mathcal{V}^t = \varnothing$. Table \ref{table-notation} demonstrates the summary of notations.

\begin{table}[]
\centering
\caption{Summary of Notations}
\small
\begin{tabular}{@{}c|l@{}}
\toprule
\textbf{Notations}   & \textbf{Description}                               \\ \midrule

$u_i \in \mathbb{N}_0$                   & categorical variables for $i$-th user            \\ \hline
$v_j \in \mathbb{N}_0$                   & categorical variables for $j$-th item            \\ \hline
$\mathcal{U}^s=\left\{u_1^s, u_2^s, \ldots\right\}$                   & User set in the source domain              \\ \hline
$\mathcal{V}^s=\left\{v_1^s, v_2^s, \ldots\right\}$                   & Item set in the source domain              \\ \hline
$\mathcal{U}^t=\left\{u_1^t, u_2^t, \ldots\right\}$                   & User set in the target domain              \\ \hline
$\mathcal{V}^t=\left\{v_1^t, v_2^t, \ldots\right\}$                   & Item set in the target domain              \\ \hline
$I_{ij}$                    & indicator function for $u_i$ and $v_j$  \\ \hline
$\mathbf{R}$                    & Interaction matrix                                 \\ \hline
$d \in \mathbb{N}$                    & The dimension of latent representation             \\ \hline
$\mathbf{U} \in \mathbb{R}^{|\mathcal{U}| \times d}$                   & latent representation Matrix for $\mathcal{U}$              \\ \hline
$\mathbf{V}\in \mathbb{R}^{|\mathcal{V}| \times d}$                   & latent representation Matrix for $\mathcal{V}$              \\ \hline
$\mathbf{u}_i \in \mathbb{R}^d$                   & latent user representation embedding for $u_i$              \\ \hline
$\mathbf{v}_i \in \mathbb{R}^d$                   & latent item representation embedding for $v_i$              \\ \hline
$\mathcal{U}^o$ & overlapping user, $\mathcal{U}^o=\mathcal{U}^s \cap \mathcal{U}^t$                                   \\ \hline
$\hat{\mathbf{u}}^s \in \mathbb{R}^d$              & overlapping user in source domain            \\ \hline
$\hat{\mathbf{u}}^t \in \mathbb{R}^d$              & overlapping user in target domain            \\ \hline
$f_U$                    & mapping function                    \\ \hline
$\theta$                & parameters of the mapping function                  \\ \hline
$\delta$                & perturbation                  \\ \hline
$k$                    & PGD step size                \\ \hline
$\rho$                    & radius of perturbation                    \\ \hline
$\alpha$                    & perturbation rate in PGD method                   \\ \hline
$\eta$                    & perturbation of FGSM attack                   \\ \hline

\end{tabular}
\label{table-notation}
\end{table}

In this study, we aim to recommend items to overlapping cold-start users $\mathcal{U}^o$ in the target domain. This problem corresponds to the real-world scenario, where the cold-start users $\mathcal{U}^o$ have no interaction with $\mathcal{V}^t$, and we leverage their interaction history with $\mathcal{V}^s$ to initialize their latent representation in the target domain. It is worth noting that our problem setting differs from other CDR tasks \cite{liu2021leveraging, sun2023remit} where auxiliary data was utilized for training.

\begin{figure*}[t]
\centerline{\includegraphics[width=0.99\textwidth]{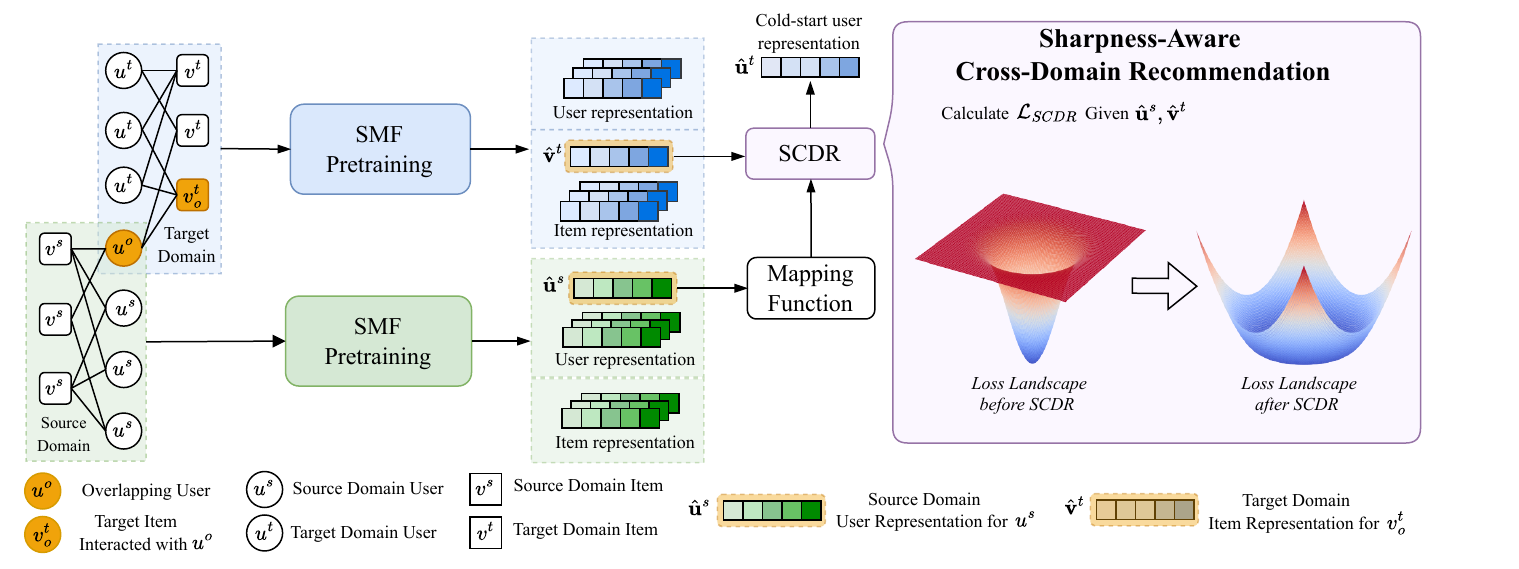}}
\caption{Illustration of our proposed Sharpness-Aware Cross-Domain Recommendation (SCDR) method. SCDR optimizes the CDR model to converge to a flatter minima through a min-max optimization approach.}
\label{fig2}
\end{figure*}

\section{Proposed Methodology}

In this section, we introduce the technical details of our proposed SCDR method. SCDR follows the paradigm of EMCDR-based method, which consists of three steps. Firstly, we pretrain the latent representations for the source domain and target domain using the matrix factorization method. Next, we train the mapping function across the two domains with Sharpness-Aware Minimization to obtain a well-generalized model for CDR. Finally, we infer the latent representation for cold-start users in the target domain by utilizing their pre-trained representation in the source domain and the mapping function. 

\subsection{Latent Representation Pre-Training}

Following the EMCDR-based methods \cite{man2017cross, zhu2022personalized}, we first pre-train the latent representation for both domains, then the mapping function is learned to connect the two representation spaces. We use the Probabilistic Matrix Factorization (MF) model \cite{mnih2007probabilistic} to map each user $u$ and item $v$ to a vector space $\mathbb{R}^d$. Let $\mathbf{U}=[\mathbf{u}_1,...,\mathbf{u}_{\left|\mathcal{U}\right|}] \in \mathbb{R}^{\left|\mathcal{U}\right| \times d}$ and $\mathbf{V}=[\mathbf{v}_1,...,\mathbf{v}_{\left|\mathcal{V}\right|}] \in \mathbb{R}^{\left|\mathcal{V}\right| \times d}$ denote the learnt representation matrices for $\mathcal{U}$ and $\mathcal{V}$, and $d$ is the dimension of latent space. In this paper, we use $\mathbf{u}_i \in \mathbb{R}^d$ and $\mathbf{v}_j \in \mathbb{R}^d$ to denote the latent representation of the user $u_i$ and item $v_j$, respectively. Based on \cite{mnih2007probabilistic}, the conditional distribution over the interaction matrix $\mathbf{R}$ is defined as follows:
\begin{equation}
p\left(\mathbf{R} \mid \mathbf{U}, \mathbf{V}, \sigma^2\right)=\prod_{i=1}^{\left|\mathcal{U}\right|} \prod_{j=1}^{\left|\mathcal{V}\right|}\left[\mathcal{N}\left(R_{i j} \mid \Braket{\mathbf{u}_i,\mathbf{v}_j}, \sigma^2\right)\right]^{I_{i j}}
\label{eq_newMF}
\end{equation}
where $\mathcal{N}\left(\mathbf{x} \mid \mu, \sigma^2\right)$ is a Gaussian distribution with mean $\mu$ and variance $\sigma^2$. $I_{ij}$ is an indicator function, which is set to 1 if the $i$-th user has an interaction history with the $j$-th item, and 0 otherwise. In Eq. (\ref{eq_newMF}), $\Braket{\mathbf{u}_i,\mathbf{v}_j}$ represents inner product of $\mathbf{u}_i$ and $\mathbf{v}_j$. The Probabilistic MF model maximizes the probability of the conditional distribution over the interaction matrix $\mathbf{R}$ by solving the following optimization problem
\begin{equation}
\begin{aligned}
\min _{\mathbf{U}, \mathbf{V}}\mathcal{L}_{\rm MF}(\mathbf{u},\mathbf{v})=\sum_i \sum_j\left\|I_{i j} \cdot\left(R_{i j}- \Braket{\mathbf{u}_i,\mathbf{v}_j}\right)\right\|_F^2
\end{aligned}
\label{eq-mf}
\end{equation}
where $\mathcal{L}_{\rm MF}$ denotes the loss function for Matrix Factorization and $\|\cdot\|_F^2$ is the Frobenius norm. The parameters of $\mathbf{U}$ and $\mathbf{V}$ can be learned by optimizing the above objective function via stochastic gradient descent. We train two MF models for the two domains, respectively. Intuitively, it learns a representation space for describing the preference for users and items. Therefore, we can recommend items to users based on their similarity in the latent representation.

\subsection{Sharpness-Aware Minimization for CDR}

Given the latent representation that had been trained in the first step, EMCDR \cite{man2017cross} learns the explicit mapping function by minimizing the following loss function
\begin{equation}
\min _{\theta} \sum_{u_i \in \mathcal{U}^o} \mathcal{L}_{\rm MSE}\left(f_U\left(\mathbf{u}_i^s ; \theta\right), \mathbf{u}_i^t\right)
\end{equation}
where $f_U(\cdot; \theta): \mathbb{R}^d \to \mathbb{R}^d$ is the mapping function, $\theta$ denotes its parameters, $\mathcal{L}_{\rm MSE}$ is Mean Square Error loss function, and $\mathbf{u}_i^s$ and $\mathbf{u}_i^t$ denote the latent representation in the {two domains corresponding to $u_i$}. EMCDR minimizes the distance between $f_U\left(\mathbf{u}_i^s ; \theta\right)$ and $\mathbf{u}_i^t$ to build an explicit connection between the two domains, to transfer the cold-start users from the source domain to the target domain. However, due to the small number of overlapping users $\mathcal{U}^o$ available for training the model, EMCDR-based methods often result in inferior recommendation performance. As we show in Figure \ref{fig:landscape} (a), the loss landscape of the EMCDR method converges to sharp minima, and a sharp loss landscape is commonly associated with poor generalization \cite{wu2020adversarial, wen2022sharpness}.

Motivated by recent advancements in \cite{foret2020sharpness}, we consider enhancing the generalization capability of CDR from the perspective of the geometric properties of the loss. Without loss of generality, we consider a family of models parameterized by $\boldsymbol{w}$ and the following optimization problem
\begin{equation}
\min _{\boldsymbol{w}} \mathcal{L}(\boldsymbol{w})+\mathcal{R}(\boldsymbol{w})
\label{eq_sharp-0}
\end{equation}
where $\mathcal{L}$ is an arbitrary loss function. and $\mathcal{R}(\boldsymbol{w})$ is a sharpness regularization term which defined as
\begin{equation}
\mathcal{R}(\boldsymbol{w})=\max _{\|\boldsymbol{\delta}\|_2 \leq \rho} \mathcal{L}(\boldsymbol{w}+\boldsymbol{\delta})-\mathcal{L}(\boldsymbol{w})
\label{eq_sharp}
\end{equation}
where $\delta$ denotes a perturbation under the constraint $\|\delta\|_2 \leq \rho$, and $\rho \in \mathbb{R}^+$ is a hyper-parameter to control the radius of perturbation. $\mathcal{R}(\boldsymbol{w})$ characterizes the geometric properties of the loss landscape, with $\mathcal{R}(\boldsymbol{w})$ taking larger values when the loss values in the vicinity of $\boldsymbol{w}$ have a larger upper bound. Note that machine learning models may converge to different local minima \cite{kawaguchi2016deep}, minimizing the empirical error while penalizing $\mathcal{R}(\boldsymbol{w})$ will make the model converge to a flatter minima from among the many possible minima. \cite{foret2020sharpness} further derive PAC-Bayesian generalization bounds as follows:

\begin{proposition}
For any $\rho > 0$, let $\mathcal{L}_{\mathcal{D}}$ be the expected loss and $\mathcal{L}_{\mathcal{S}}$ be the training loss, where the training set S is drawn from data distribution D with i.i.d condition, then
\begin{equation}
\mathcal{L}_{\mathcal{D}}(\boldsymbol{w}) \leq \max _{\|\boldsymbol{\delta}\|_2 \leq \rho} \mathcal{L}_{\mathcal{S}}(\boldsymbol{w}+\boldsymbol{\delta})+h\left(\| \boldsymbol{w}  \|_2^2 / \rho^2\right)
\label{pac}
\end{equation}
where $h: \mathbb{R}_{+} \rightarrow \mathbb{R}_{+}$is a strictly increasing function.
\end{proposition}
Note that the Eq (\ref{eq_sharp-0}) can be rewritten as $\max _{\|\boldsymbol{\delta}\|_2 \leq \rho} \mathcal{L}_{\mathcal{S}}(\boldsymbol{w}+\boldsymbol{\delta})$, which lies in the right hand side of Eq. (\ref{pac}). Eq. (\ref{pac}) provides a theoretical connection between the model's generalization ability and the loss landscape, indicating that optimizing the loss landscape sharpness leads to better generalization.

Now, let us consider the sharpness-aware minimization approach for a CDR scenario to address the aforementioned issues caused by the limited overlapping training data. Specifically, we consider the learnable parameters $\boldsymbol{w}$ to be the user representation for overlapping users and the mapping function in cross-domain recommendation. 
Let $( \hat{\mathbf{\mathbf{u}}}^s, \hat{\mathbf{\mathbf{v}}}^t) \sim D$ denotes the latent representations pairs corresponding to $\hat{u} \in \mathcal{U}^o$, where $\hat{\mathbf{\mathbf{u}}}^s\in \mathbb{R}^d$ represents the user representation for $\hat{u}$ in the source domain, $\hat{\mathbf{\mathbf{v}}}^t \in \mathbb{R}^d$ is the item representation for $\hat{v}$ in the target domain, and $D$ denotes the distribution of latent representations pairs $( \hat{\mathbf{\mathbf{u}}}^s, \hat{\mathbf{\mathbf{v}}}^t)$. $\theta$ denotes the parameters of the mapping function. Inspired by Sharpness-aware minimization, we aim to optimize the following problem
\begin{equation}
\sup _{\|\boldsymbol{\delta}\|_2 \leq \rho} \mathbb{E}_{( \hat{\mathbf{u}}^s, \hat{\mathbf{v}}^t) \sim D} \mathcal{L}_{\rm MF}(f_{U}(\hat{\mathbf{u}}^s+\boldsymbol{\delta}; \theta), \hat{\mathbf{v}}^t)
 \label{eq_at}
\end{equation}
where $\mathcal{L}_{\rm MF}$ is the Matrix Factorization loss defined in Eq. (\ref{eq-mf}). Eq. (\ref{eq_at}) induces perturbations on $\mathbf{u}^o$, enabling the model to converge to a flat minimum where the loss landscape is relatively smooth for the overlapping users. From the perspective of the CDR task, Eq. (\ref{eq_at}) encourages the representation space of the neighborhood of overlapping users to become flatter, allowing more user representation to maintain better generalization capability after being mapped to the target domain, especially when the training data for overlapping users is limited. Note that accurately optimizing Eq. (\ref{eq-outer}) is an NP-hard problem \cite{sinha2017certifying}, therefore we approximate the optmization problem in Eq. (\ref{eq-inner}) as a bi-level optimization problem:
\begin{equation}
\min _{\theta, \hat{\mathbf{u}}^s} \mathcal{L}_{SCDR}(\theta, \hat{\mathbf{u}}^s)
\label{eq-outer}
\end{equation}
where
\begin{equation}
\mathcal{L}_{SCDR}(\theta, \hat{\mathbf{u}}^s) \triangleq \max _{\|\boldsymbol{\delta}\|_2 \leq \rho} \mathcal{L}_{\rm MF}((f_U(\hat{\mathbf{u}}^s)+\boldsymbol{\delta}; \theta), \hat{\mathbf{v}}^t)
\label{eq-inner}
\end{equation}
Eq. (\ref{eq-outer}) is the outer optimization problem and Eq. (\ref{eq-inner}) is the inner optimization problem. We denote the proposed SCDR loss as $\mathcal{L}_{SCDR}$. Similarly, we apply sharpness-aware minimization on the pre-training stage to enhance the generalization ability of user representations. We replace Eq. (\ref{eq-mf}) with
\begin{equation}
\begin{aligned}
\mathcal{L}_{\rm SMF}(\mathbf{u},\mathbf{v}) \triangleq \max _{\|\boldsymbol{\delta}\|_2 \leq \rho} \mathcal{L}_{\rm MF}(\mathbf{u} +\boldsymbol{\delta} ,\mathbf{v})
\end{aligned}
\label{eq-smf}
\end{equation}
We refer to $\mathcal{L}_{\rm SMF}$ as the Sharpness-Aware Matrix Factorization (SMF) loss for user representations in both domains. The motivation behind SMF is to further flatten the user representation loss landscape during the pre-training phase.

The key to training the SCDR model lies in 1: how to find the perturbation $\boldsymbol{\delta}$ that maximizes the inner optimization, and 2: how to optimize the outer optimization. In the next section, we will explain in detail how to optimize the proposed method.

%start from here

\subsection{Optimization for SCDR}

To approximate the optimization problem in Eq. (\ref{eq_at}), the key is to training SCDR is to find a perturbation $\boldsymbol{\delta}$ to solve the inner objective $\mathcal{L}_{SCDR}$. Let $\boldsymbol{\delta}^*$
denote the approximated perturbation that solves the inner optimization, \cite{foret2020sharpness} apply first-order Taylor expansion of $\mathcal{L}(\boldsymbol{w} + \boldsymbol{\delta})$ to find $\boldsymbol{\delta}^*$
\begin{equation}
\begin{aligned}
\boldsymbol{\delta}^*  \approx \underset{\|\boldsymbol{\delta}\|_2 \leq \rho}{\arg \max } \mathcal{L} (\boldsymbol{w})+\boldsymbol{\delta}^T \nabla_{\boldsymbol{w}} L(\boldsymbol{w})  =\underset{\|\boldsymbol{\delta} \|_2 \leq \rho}{\arg \max } \boldsymbol{\delta}^T \nabla_{\boldsymbol{w}} L(\boldsymbol{w}) .
\end{aligned}
\label{eq-taylor}
\end{equation}
Eq. (\ref{eq-taylor}) approximates $\boldsymbol{\delta}^*$ through one-step gradient ascent, and \cite{foret2020sharpness} found that using Eq. (\ref{eq-taylor}) to find the perturbation can achieve sufficiently good performance on image dataset. In this work, we discover through experiments that finding $\boldsymbol{\delta}$ through multi-step gradient ascent can endow the CDR model with better robustness against adversarial attacks. Specifically, we use Projected Gradient Descent (PGD) \cite{madry2017towards} to find a better approximated $\boldsymbol{\delta}$ to maximize $\mathcal{L}_{SCDR}$. PGD finds the value that maximizes the loss function by iteratively calculating the gradient ascent direction. In particular, let $\hat{\mathbf{u}}_k^s \in \mathbb{R}^d$ denotes the perturbed overlapping users representation at the $k$-th steps, PGD iteratively perturb sample $\hat{\mathbf{u}}^s$ as follows:
\begin{equation}
\hat{\mathbf{u}}_{k+1}^{s}=\Pi_{\varepsilon}\left(\hat{\mathbf{u}}_k^{s}+\alpha \operatorname{sgn}\left(\nabla_{\hat{\mathbf{u}}^s} \mathcal{L}_{MF}( \hat{\mathbf{u}}^{s}, \hat{\mathbf{v}}^t)\right)\right)
\label{eq-adv1}
\end{equation}
where $\alpha$ denotes the attack learning rate, $k \in \mathbb{N}_0$ indicates the number of step, and $\hat{\mathbf{u}}_0^s = \hat{\mathbf{u}}^s$ denotes the initial representation that has not been perturbed. $\Pi_{\varepsilon}(\cdot)$ is the project function defined as
\begin{equation}
\Pi_{\rho}(\hat{\mathbf{u}}^{s}_k)= \begin{cases}\hat{\mathbf{u}}^s_0+\frac{\hat{\mathbf{u}}^{s}_k-\hat{\mathbf{u}}^s_0}{\left\|\hat{\mathbf{u}}^{s}_k-\hat{\mathbf{u}}^s_0\right\|_2} \rho & \text { if }\left\|\hat{\mathbf{u}}^{s}_k-\hat{\mathbf{u}}^s_0\right\|_2>\rho \\ \hat{\mathbf{u}}^{s}_k & \text { otherwise }\end{cases}
\label{eq-adv2}
\end{equation}
Eq. (\ref{eq-adv2}) ensures that $\hat{\mathbf{u}}^{s}_k$ stays within the $\ell_2$ norm ball of $\hat{\mathbf{u}}^{s}_0$. We calculate the perturbed $\hat{\mathbf{u}}^{s}$ through Eq. (\ref{eq-adv1}) and Eq. (\ref{eq-adv2}). 
Calculating $\boldsymbol{\delta}$ with the PGD ensures that the perturbed parameters remain within the $\ell_p$ norm ball range of the original parameters, which satisfies the definition of sharpness-aware minimization in Eq. (\ref{eq-inner}). 

We train the proposed SCDR model by solving the bi-level optimization defined in Eq. (\ref{eq-inner}) and Eq. (\ref{eq-outer}) via stochastic gradient descent. The gradient of the mapping function $\nabla_{\theta} \mathcal{L}_{SCDR}(\theta)$ can be obtained through the standard chain rule. However, note that the perturbation $
\boldsymbol{\delta}^*$ is a function of $\hat{\mathbf{u}}^{s}$, the gradient $\nabla_{\hat{\mathbf{u}}^{s}} \mathcal{L}_{SCDR}(\hat{\mathbf{u}}^s)$  will involve the computation of the Hessian of $\mathcal{L}_{MF}$. Let $\boldsymbol{\delta}^*(\hat{\mathbf{u}}^{s})$ denote the perturbation that maximizes Eq. (\ref{eq-inner}), we have

\begin{equation}
\begin{aligned}
\nabla_{\hat{\mathbf{u}}^{s}} \mathcal{L}_{SCDR}(\hat{\mathbf{u}}^s) & \approx \nabla_{\hat{\mathbf{u}}^{s}} \mathcal{L}_{MF}(\hat{\mathbf{u}}^{s}+\boldsymbol{\delta}^*(\hat{\mathbf{u}}^{s})) \\
& =\left.\frac{d(\hat{\mathbf{u}}^{s}+\boldsymbol{\delta}^*(\hat{\mathbf{u}}^{s}))}{d \hat{\mathbf{u}}^{s}} \nabla_{\hat{\mathbf{u}}^{s}} \mathcal{L}_{MF}(\hat{\mathbf{u}}^s)\right|_{\hat{\mathbf{u}}^{s}+\boldsymbol{\delta}^*(\hat{\mathbf{u}}^{s})} \\
& =\left.\nabla_{\hat{\mathbf{u}}^{s}} \mathcal{L}_{MF}(\hat{\mathbf{u}}^s)\right|_{\hat{\mathbf{u}}^{s}+\boldsymbol{\delta}^*(\hat{\mathbf{u}}^{s})} \\
& +\left.\frac{d \boldsymbol{\delta}^*(\hat{\mathbf{u}}^{s})}{d \hat{\mathbf{u}}^{s}} \nabla_{\hat{\mathbf{u}}^{s}} \mathcal{L}_{MF}(\hat{\mathbf{u}}^s))\right|_{\hat{\mathbf{u}}^{s}+\boldsymbol{\delta}^*(\hat{\mathbf{u}}^{s})}
\end{aligned}
\end{equation}

However, note that the gradient $\nabla_{\hat{\mathbf{u}}^{s}} \mathcal{L}_{SCDR}(\hat{\mathbf{u}}^s)  \approx \nabla_{\hat{\mathbf{u}}^{s}} \mathcal{L}_{MF}(\hat{\mathbf{u}}^{s}+\boldsymbol{\delta}^*(\hat{\mathbf{u}}^{s}))$ involves the computation for $\frac{d \boldsymbol{\delta}^*\left(\hat{\mathbf{u}}^s\right)}{d \hat{\mathbf{u}}^s}$, which depends on the Hessian of $\mathcal{L}_{MF}$ and can be computationally expensive. Following \cite{foret2020sharpness}, we drop the Hessian second-order terms, obtaining gradient approximation as
\begin{equation}
\nabla_{\hat{\mathbf{u}}^{s}} \mathcal{L}_{SCDR}(\hat{\mathbf{u}}^s) \approx \left.\nabla_{\hat{\mathbf{u}}^{s}} \mathcal{L}_{MF}(\hat{\mathbf{u}}^s)\right|_{\hat{\mathbf{u}}^{s}+\boldsymbol{\delta}^*(\hat{\mathbf{u}}^{s})}
\label{eq-gradient}
\end{equation}
with $\boldsymbol{\delta}^*(\hat{\mathbf{u}}^{s})$ computed by PGD method, and the approximation of Eq. (\ref{eq-gradient}) can be computed via automatic differentiation. Similarly, we have $\nabla_{\mathbf{u}} \mathcal{L}_{SMF}(\mathbf{u}) \approx \left.\nabla_{\mathbf{u} \mathcal{L}_{MF}(\mathbf{u})}\right|_{\mathbf{u}+\boldsymbol{\delta}^*(\mathbf{u})}$. To this end, we can train the proposed SCDR via standard stochastic gradient descent.

\subsection{Cold-Start Recommendation Stage}

After training $\hat{\mathbf{u}}^s$ and the mapping function $f_U$ in the above stages, we can now make recommendations for cold-start users. Given the overlapping user $\mathcal{U}^o$, we split a portion to train the function $f_U$, and then we can make cold-start recommendations for the remaining test samples using
\begin{equation}
\hat{\mathbf{u}}^t = f_U(\hat{\mathbf{u}}^s; \theta)
\label{eq-final}
\end{equation}
Eq. (\ref{eq-final}) provides an initialization for cold-start users in the target domain even $\hat{\mathbf{u}}^t$ have no interaction with $\mathcal{U}^t$. After obtaining $\hat{\mathbf{u}}^t$, we recommend items to users based on the distance between $\hat{\mathbf{u}}^t$ and the target domain item representations. The procedure of SCDR can be found at Algorithm \ref{alg-SCDR}.

\begin{algorithm} [t] 
    \caption{Procedure of the proposed SCDR method}\label{alg}
    \flushleft{\hspace*{0.0in}\textbf{Input}: The users sets, item sets and interaction matrices of each domain $\mathcal{U}^s, \mathcal{U}^t, \mathcal{V}^s, \mathcal{V}^t$, $\mathcal{U}^o$, $\mathbf{R}^s, \mathbf{R}^t$
    
    \hspace*{0.00in}\textbf{Output}: Recommend items to cold-start users 
    
    \hspace*{0.02in}\textbf{(1) SMF PreTraining}:
    
    \hspace*{0.1in}1: Learning $\left\{\mathbf{U}^s, \mathbf{V}^s\right\}$ for source domain via Eq. (\ref{eq-smf})
        
    \hspace*{0.1in}2: Learning $\left\{\mathbf{U}^t, \mathbf{V}^t\right\}$ for target domain via Eq. (\ref{eq-smf})

    \hspace*{0.1in}3: Obtaining  $( \hat{\mathbf{\mathbf{u}}}^s, \hat{\mathbf{\mathbf{v}}}^t)$ pairs for the next step
        
    \hspace*{0.02in}\textbf{(2) Training SCDR }:

    \hspace*{0.1in}4: Training the mapping function $f_U$ and train the latent 
    \hspace*{0.1in} representation of $\hat{\mathbf{u}}^s$ by solving Eq. (\ref{eq-outer})

    \hspace*{0.02in}\textbf{(3) Cold-Start Recommendation Stage}:
    
    \hspace*{0.1in}5. Obtain cold-start user presentation in the target domain $\hat{\mathbf{u}}^t$ by using Eq. (\ref{eq-final})
    }
    \label{alg-SCDR}

\end{algorithm}

\section{Experiment}

In this section, we perform a series of experiments on CDR tasks to validate the effectiveness of our method. As our method is a EMCDR-based approach, we primarily compare it with other EMCDR-based methods. We present a series of experiments and analysis to answer the following research questions:

\begin{itemize}
    \item \textbf{RQ1}: How does SCDR perform in cold-start recommendation tasks compared to existing CDR methods?
    \item \textbf{RQ2}: What impact does SCDR have on the loss landscape?
    \item \textbf{RQ3}: What is the effect of hyperparameters in SCDR?
    \item \textbf{RQ4}: Can SCDR improve adversarial robustness?
    \item \textbf{RQ5}: Can SCDR be integrated with different recommendation models?
    \item \textbf{RQ6}: How do the sharpness-aware minimization module contribute in the performance of SCDR?
\end{itemize}

\begin{table*}[]
\centering
\caption{Experimental results on Amazon CDR Scenarios. Best results are in boldface.}
\begin{tabular}{@{}c|c|c|cc|ccccc|cc@{}}
\toprule
                       & $\beta$               & \textbf{metric} & \textbf{TGT} & \textbf{CMF} & \textbf{CLCDR} & \textbf{SSCDR} & \textbf{EMCDR} & \textbf{PTUPCDR} & \textbf{TMCDR} & \textbf{SCDR}   & \textbf{improve} \\ \midrule
\multirow{6}{*}{Scenario 1} & \multirow{2}{*}{20\%} & \textit{MAE}    & 4.4016       & 1.4948       & 1.2109         & 1.3859         & 1.3406         & 1.1147           & 1.0536         & \textbf{0.9658} & 8.33\%           \\
                       &                       & \textit{RMSE}   & 5.1023       & 1.9873       & 1.4984         & 1.6521         & 1.6450         & 1.4447           & 1.4164         & \textbf{1.2649} & 10.70\%           \\ \cmidrule(l){2-12} 
                       & \multirow{2}{*}{50\%} & \textit{MAE}    & 4.4497       & 1.5790       & 1.5429         & 1.6429         & 1.6919         & 1.3111           & 1.2026         & \textbf{0.9729} & 19.10\%          \\
                       &                       & \textit{RMSE}   & 5.1461       & 2.0991       & 1.8577         & 1.9928         & 2.0555         & 1.7719           & 1.6512         & \textbf{1.2731} & 22.90\%          \\ \cmidrule(l){2-12} 
                       & \multirow{2}{*}{80\%} & \textit{MAE}    & 4.5021       & 2.0092       & 2.1090         & 2.1245         & 2.2154         & 1.6733           & 1.5872         & \textbf{1.0127} & 36.20\%          \\
                       &                       & \textit{RMSE}   & 5.1860       & 2.4988       & 2.4301         & 2.5928         & 2.6041         & 2.3059           & 2.1870         & \textbf{1.3316} & 39.11\%          \\ \midrule
\multirow{6}{*}{Scenario 2} & \multirow{2}{*}{20\%} & \textit{MAE}    & 4.2125       & 1.4152       & 1.0821         & 1.3470         & 1.1190         & 1.0578           & 0.9228         & \textbf{0.9133} & 1.03\%           \\
                       &                       & \textit{RMSE}   & 4.7785       & 1.8507       & 1.3543         & 1.5173         & 1.4144         & 1.3630           & 1.2081         & \textbf{1.1985} & 0.79\%           \\ \cmidrule(l){2-12} 
                       & \multirow{2}{*}{50\%} & \textit{MAE}    & 4.2127       & 1.4966       & 1.1609         & 1.2229         & 1.1934         & 1.1050           & 0.9731         & \textbf{0.9543} & 1.93\%           \\
                       &                       & \textit{RMSE}   & 4.7818       & 1.9561       & 1.5212         & 1.5921         & 1.5042         & 1.4412           & 1.3107         & \textbf{1.2621} & 3.71\%           \\ \cmidrule(l){2-12} 
                       & \multirow{2}{*}{80\%} & \textit{MAE}    & 4.3286       & 2.1157       & 1.2944         & 1.4616         & 1.3100         & 1.2007           & 1.0867         & \textbf{1.0390} & 4.39\%           \\
                       &                       & \textit{RMSE}   & 4.8195       & 2.6315       & 1.6982         & 1.8523         & 1.6668         & 1.6008           & 1.4854         & \textbf{1.4215} & 4.30\%           \\ \midrule
\multirow{6}{*}{Scenario 3} & \multirow{2}{*}{20\%} & \textit{MAE}    & 4.4928       & 1.7502       & 1.4520         & 1.6569         & 1.6170         & 1.2494           & 1.1139         & \textbf{1.0223} & 8.23\%           \\
                       &                       & \textit{RMSE}   & 5.1418       & 2.2527       & 1.7492         & 1.9121         & 1.9222         & 1.6473           & 1.5343         & \textbf{1.3937} & 9.16\%           \\ \cmidrule(l){2-12} 
                       & \multirow{2}{*}{50\%} & \textit{MAE}    & 4.4850       & 1.8613       & 1.8532         & 2.3481         & 1.9839         & 1.4171           & 1.2799         & \textbf{1.1603} & 9.34\%           \\
                       &                       & \textit{RMSE}   & 5.1798       & 1.4328       & 2.1580         & 2.7174         & 2.3232         & 1.9224           & 1.7957         & \textbf{1.6036} & 10.69\%          \\ \cmidrule(l){2-12} 
                       & \multirow{2}{*}{80\%} & \textit{MAE}    & 5.5332       & 2.5360       & 2.2164         & 2.0417         & 2.2694         & 1.6388           & 1.5599         & \textbf{1.4026} & 10.08\%          \\
                       &                       & \textit{RMSE}   & 5.2148       & 3.2102       & 2.5957         & 2.2411         & 2.6335         & 2.2423           & 2.2182         & \textbf{1.9738} & 11.02\%          \\ \bottomrule
\end{tabular}
\label{table-main}
\end{table*}

\subsection{Experimental Settings}
\textbf{Datasets and Evaluation Metrics:} In line with the EMCDR-based methods \cite{man2017cross,zhu2022personalized}, we use the Amazon 5-scores dataset \cite{he2016ups} as our benchmark. Following \cite{zhu2022personalized}, we choose three tasks from the Amazon 5-scores dataset (\textit{Movies}, \textit{Music}, and \textit{Books}), then we defined three distinct cross-domain recommendation scenarios 1) Movies$\rightarrow$Music, 2) Books$\rightarrow$Movies, and 3) Books$\rightarrow$Music. In each scenario, the source domain contains more interaction data compared to the target domain. We randomly excluded all overlapping users in the target domain, using them as the test data. For every scenario, we designated proportions for cold-start users, labelled as $\beta$, at 20\%, 50\%, and 80\%. The remaining overlapping users serve to train the mapping function. Given that the dataset predictions are rating scores that range from 0 to 5, we employed the Mean Absolute Error (MAE) and Root Mean Square Error (RMSE) as evaluation metrics:
\begin{itemize}
    \item \textit{MAE}: measures the difference between the rating predicted by the model and the actual rating of the user.
    \begin{equation}
M A E=\frac{1}{|R|} \sum_i \sum_j\left|I_{i j} \cdot\left(R_{i j}- \Braket{\mathbf{u}_i,\mathbf{v}_j}\right)\right|
\end{equation}
\item RMSE: is the quadratic mean of the differences between the rating predicted by the model and the actual rating of the user. %the rating predicted by the model and the ground truth.
\begin{equation}
R M S E=\sqrt{\frac{1}{|R|} \sum_i \sum_j\left\|I_{i j} \cdot\left(R_{i j}- \Braket{\mathbf{u}_i,\mathbf{v}_j}\right)\right\|^2_F}
\end{equation}

\end{itemize}

The SCDR method capitalizes on the data-rich source domain to address the cold-start recommendation problem in the target domain. Statistics of each scenario are provided in Table \ref{table-datasets}.

\begin{table}[]
\centering
\caption{Statistics of the CDR scenarios. Top row denotes the source domain and bottom row is the target domain. \textit{\#Overlap} represents the number of overlapping users, \textit{ratio} represents the proportion of \textit{\#Overlap} in the total number of users.}%\textit{\#Overlap} is the number of overlapping users.}
\begin{tabular}{@{}ccccc@{}}
\toprule
                            & \textbf{} & \textbf{\#Users} & \textbf{\#Items} & \textbf{\#Overlap (ratio)} \\ \midrule
\multirow{2}{*}{Scenario 1} & Moive     & 123,960          & 50,052           & \multirow{2}{*}{18,031 (9.95\%)} \\
                            & Music     & 72,258           & 64,443           &                                  \\ \midrule
\multirow{2}{*}{Scenario 2} & Books     & 603,668          & 367,982          & \multirow{2}{*}{37,388 (5.42\%)} \\
                            & Moive     & 123,960          & 50,052           &                                  \\ \midrule
\multirow{2}{*}{Scenario 3} & Books     & 603,668          & 367,982          & \multirow{2}{*}{16,738 (2.53\%)} \\
                            & Music     & 72,258           & 64,443           &                                  \\ \bottomrule
\end{tabular}
\label{table-datasets}
\end{table}

\textbf{Baselines:} As SCDR is a EMCDR-based approach, we primarily compare it with other EMCDR-based methods. In particular, we choose the following methods as baselines: 

\begin{enumerate}
    \item MF \cite{mnih2007probabilistic} denotes training a Matrix Factorization model only on the target domain, which serves as the baseline for non-CDR methods in our comparative experiments;
    \item CMF \cite{singh2008relational} is a collective matrix factorization model that trains the users and items representation from the source and target domains simultaneously;
    \item CLCDR \cite{chen2022clcdr} is a contrastive-based CDR method for cold-start recommendation; 
    \item SSCDR \cite{kang2019semi} leverages the cross-domain relationships between users and items to construct additional supervisory information;
    \item EMCDR \cite{man2017cross} is the first work proposed to train a mapping function to solve the cold-start problem in CDR tasks; 
    \item PTUPCDR \cite{zhu2022personalized} utilizes a meta network to train a personalized mapping function; 
    \item TMCDR \cite{zhu2021transfer} proposes a transfer-meta framework for EMCDR-based CDR tasks, it learns a task-oriented network to solve the cold-start problem.
\end{enumerate}

Table \ref{table-method} demonstrate the taxonomy of the compared methods.

\begin{table}[]
\centering
\caption{Taxonomy of the compared methods}

\begin{tabular}{@{}ccc@{}}
\toprule
Method  & CDR Methods & EMCDR-based CDR Methods \\ \midrule
MF      & $\times$                    & $\times$                  \\
CMF     & $\checkmark$                & $\times$                  \\
SSCDR   & $\checkmark$                & $\checkmark$              \\
CLCDR   & $\checkmark$                & $\checkmark$              \\
EMCDR   & $\checkmark$                & $\checkmark$              \\
PTUPCDR   & $\checkmark$                & $\checkmark$              \\
TMCDR   & $\checkmark$                & $\checkmark$              \\
SCDR    & $\checkmark$                & $\checkmark$              \\ \bottomrule

\end{tabular}
\label{table-method}
\end{table}

\textbf{Implementation Details:} In this section, we detail our implementation. We employ stochastic gradient descent (SGD) to train the proposed SCDR model, which encompasses both the Probabilistic MF model and the mapping function. The learning rate is set at 0.01, while the dimension of the latent representation for the MF model is 10. The mapping function is designed as a two-layer Multi-Layer Perceptron (MLP), with a hidden dimension of 50. We adopt the tan-sigmoid activation function in the MLP to ensure a smooth mapping function. For training, the batch size is determined as 256 for all Scenarios. In the PGD training process to optimize $\mathcal{L}_{SCDR}$, we set the values of $k$ and radius $\rho$ to 5 and 5, respectively.

\subsection{Recommendation Performance (RQ1)}

We present our main experimental results in this section. For each experiment, we run trials with three random seeds and report the mean values. We adopt MAE and RMSE as our evaluation metrics. As shown in Table \ref{table-main}, our method surpasses other baselines in all tested scenarios. Notably, Matrix Factorization (MF), a non-CDR method, exhibits the poorest performance. This underscores the challenges conventional Collaborative Filtering methods face when addressing the cold-start user recommendation problem. Among the baseline methods compared, SSCDR, EMCDR, CLCDR, TMCDR and PTUPCDR utilize a EMCDR-based method, with TMCDR emerging as the top-performing method. Nevertheless, our method significantly exceeds the performance of TMCDR. In summary, our experimental findings validate the efficacy of our proposed SCDR method, emphasizing its promise to address the challenges of the cold-start problem.

Table \ref{table-main} reveals the relationship between the performance improvement of SCDR and the sparsity of overlapping users. Note that: (1) As a larger $\beta$ implies that fewer overlapping users can be used to train the mapping function, the improvement of SCDR increases with the increase of $\beta$, and (2) SCDR achieves the maximum improvement when the number of overlapping users training data is minimal (Scenario 1, $\beta=80\%$). The above two observations verify that SCDR can effectively alleviate the poor generalization problem caused by a small number of overlapping users.

\begin{figure*}[!t]
\centering
\begin{tabular}{l}
\hspace{-0.3cm}
\subfigure[Task1, EMCDR]{\includegraphics[width=0.21\linewidth]{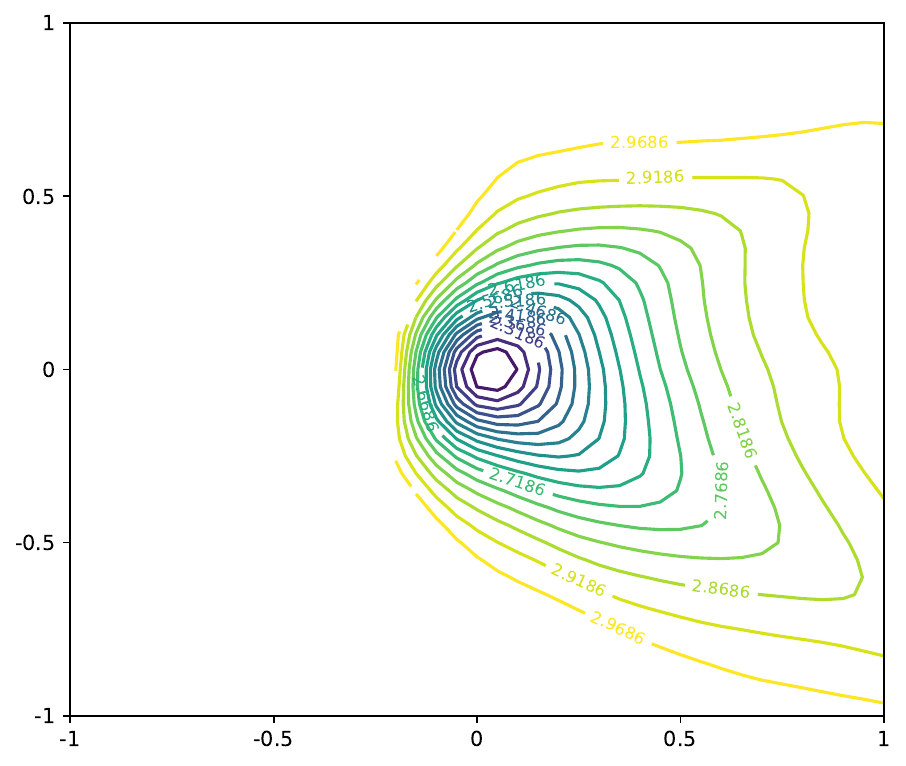}}
\subfigure[Task1, PTUPCDR]{\includegraphics[width=0.21\linewidth]{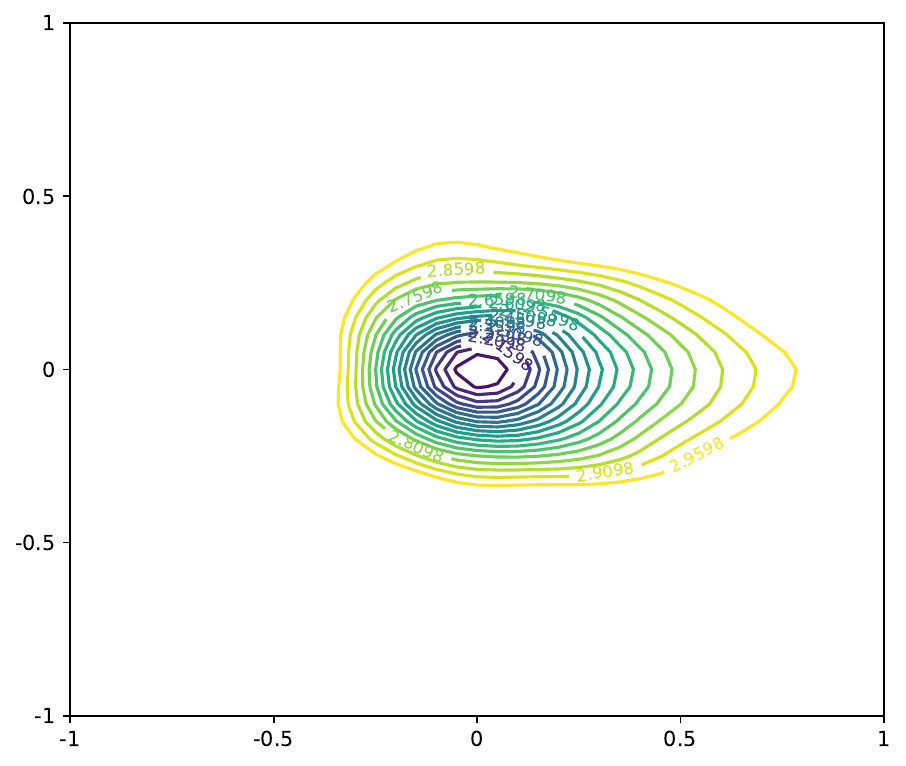}}
\subfigure[Task1, TMCDR]{\includegraphics[width=0.21\linewidth]{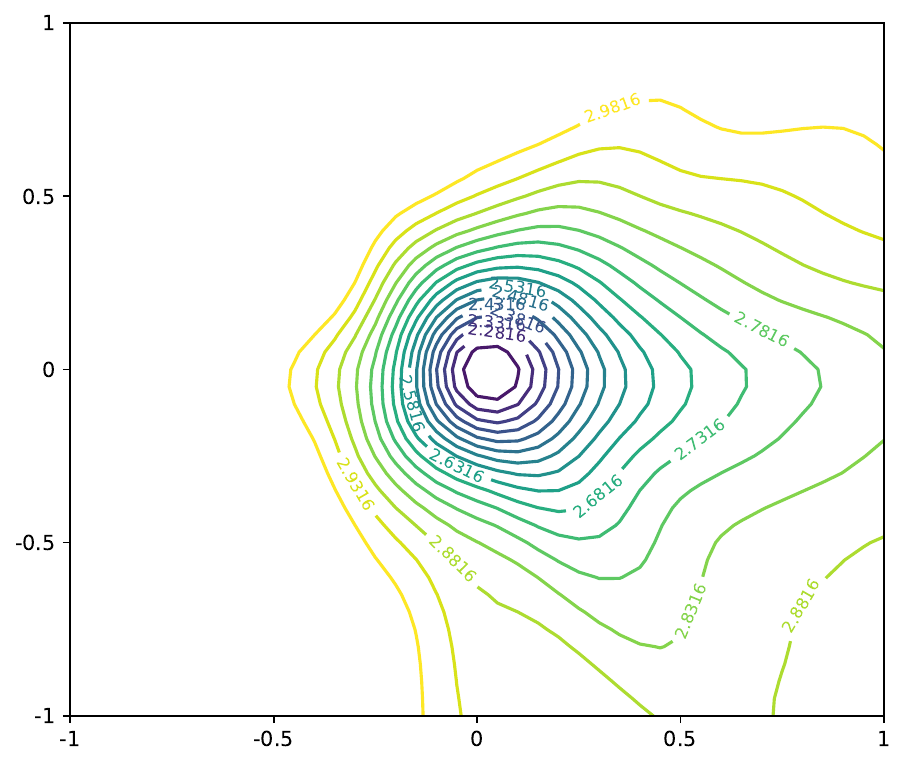}}
\subfigure[Task1, SCDR]{\includegraphics[width=0.21\linewidth]{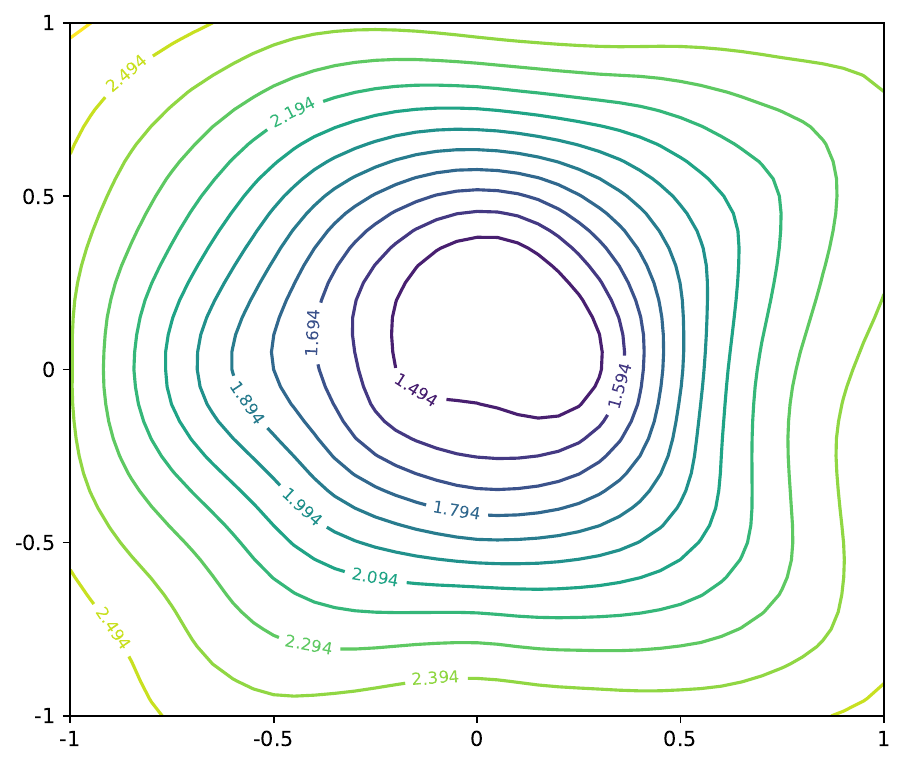}}\\
\hspace{-0.3cm}
\subfigure[Task2, EMCDR]{\includegraphics[width=0.21\linewidth]{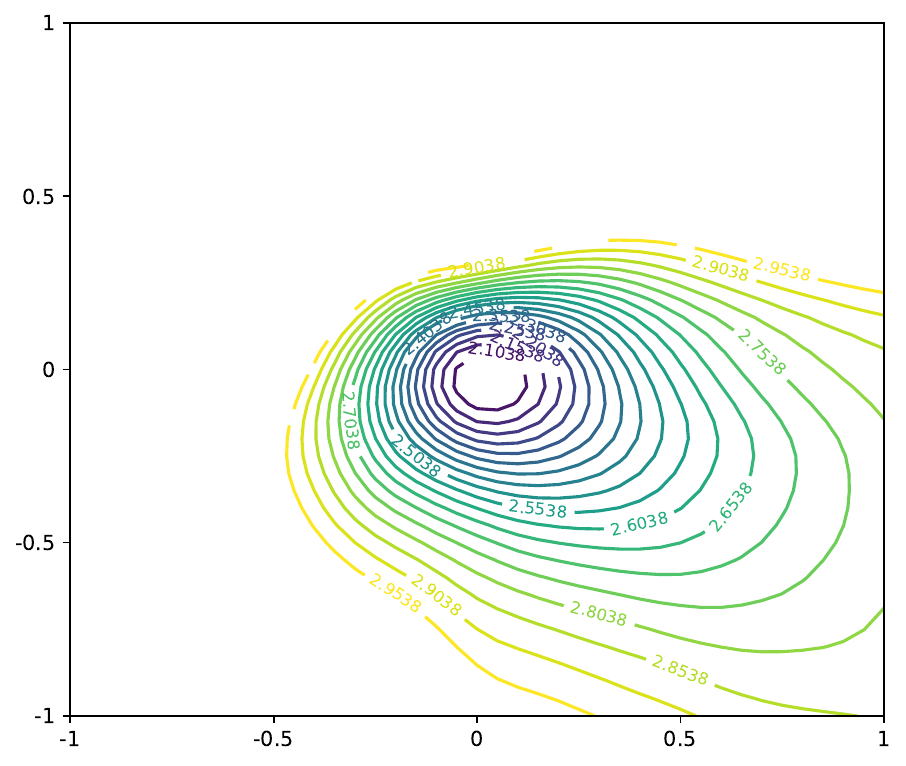}}
\subfigure[Task2, PTUPCDR]{\includegraphics[width=0.21\linewidth]{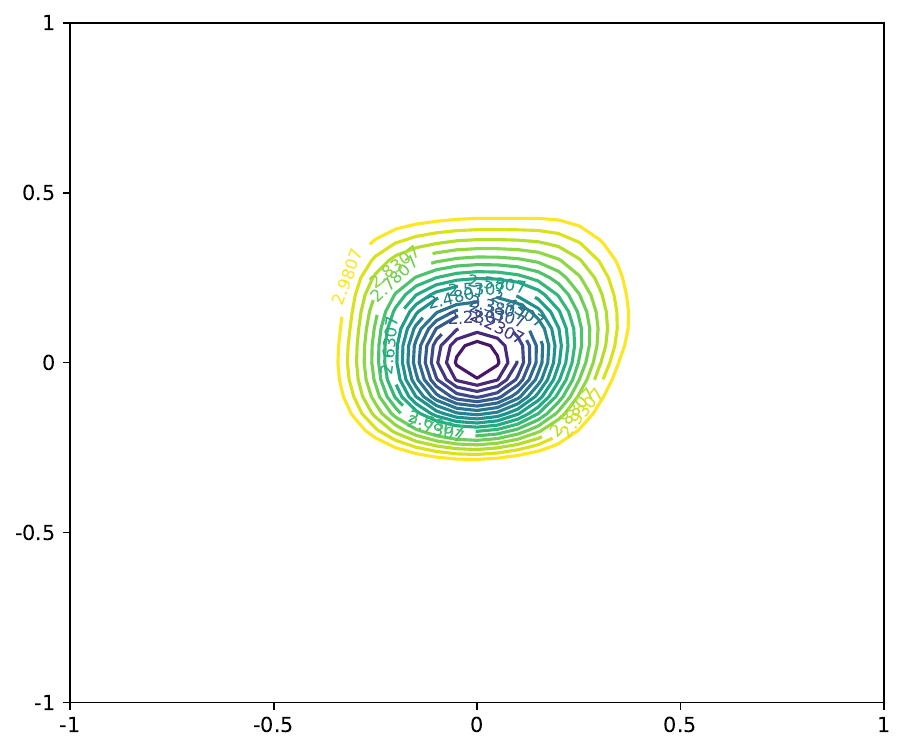}}
\subfigure[Task2, TMCDR]{\includegraphics[width=0.21\linewidth]{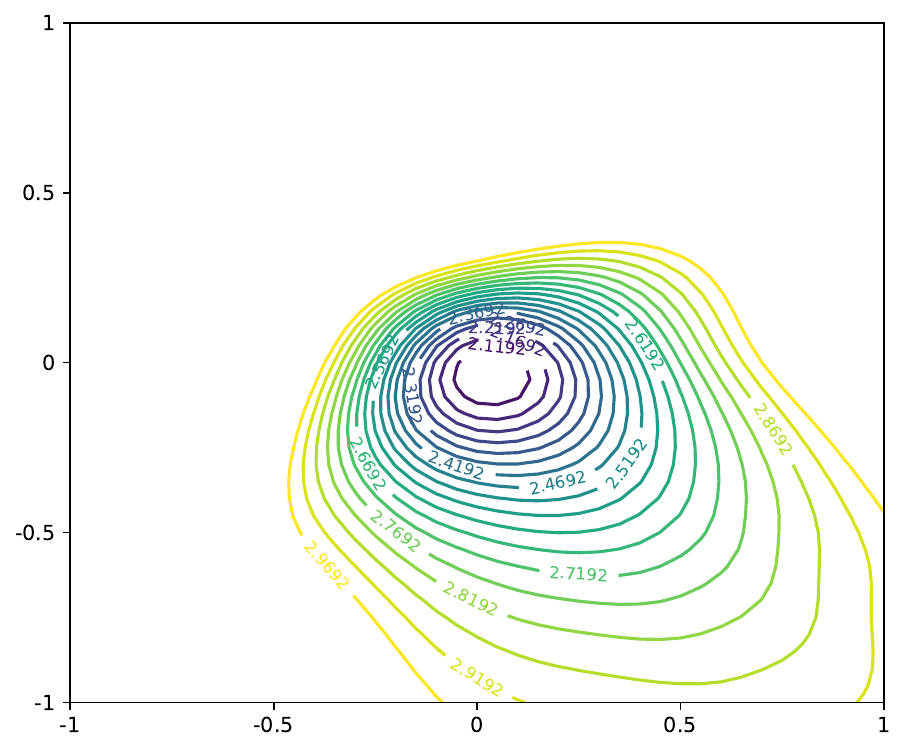}}
\subfigure[Task2, SCDR]{\includegraphics[width=0.21\linewidth]{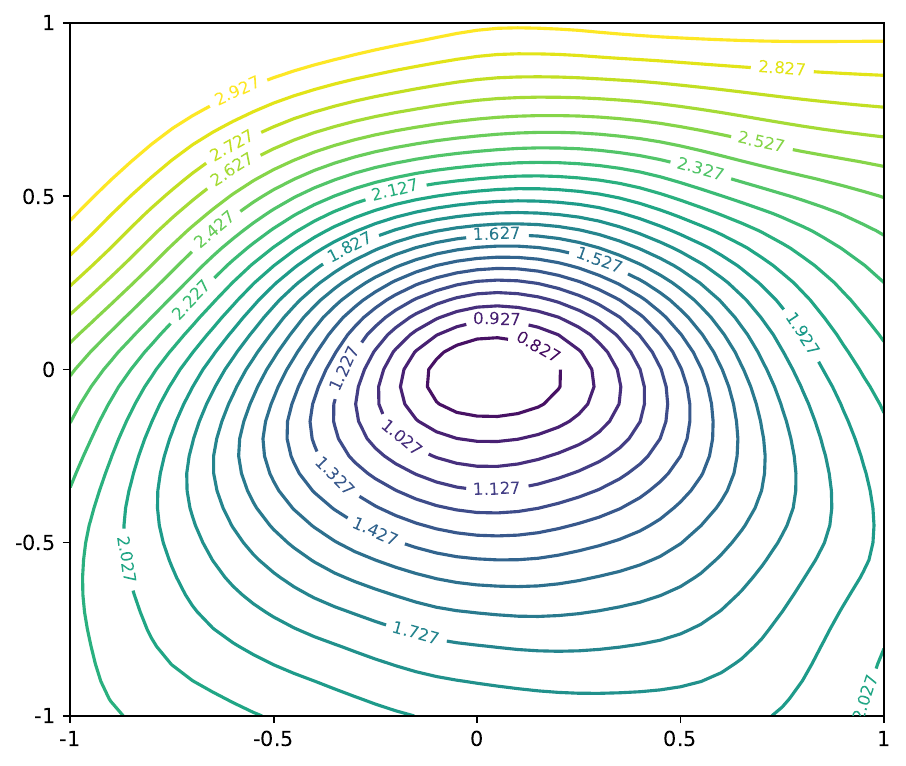}}\\
\hspace{-0.3cm}
\subfigure[Task3, EMCDR]{\includegraphics[width=0.21\linewidth]{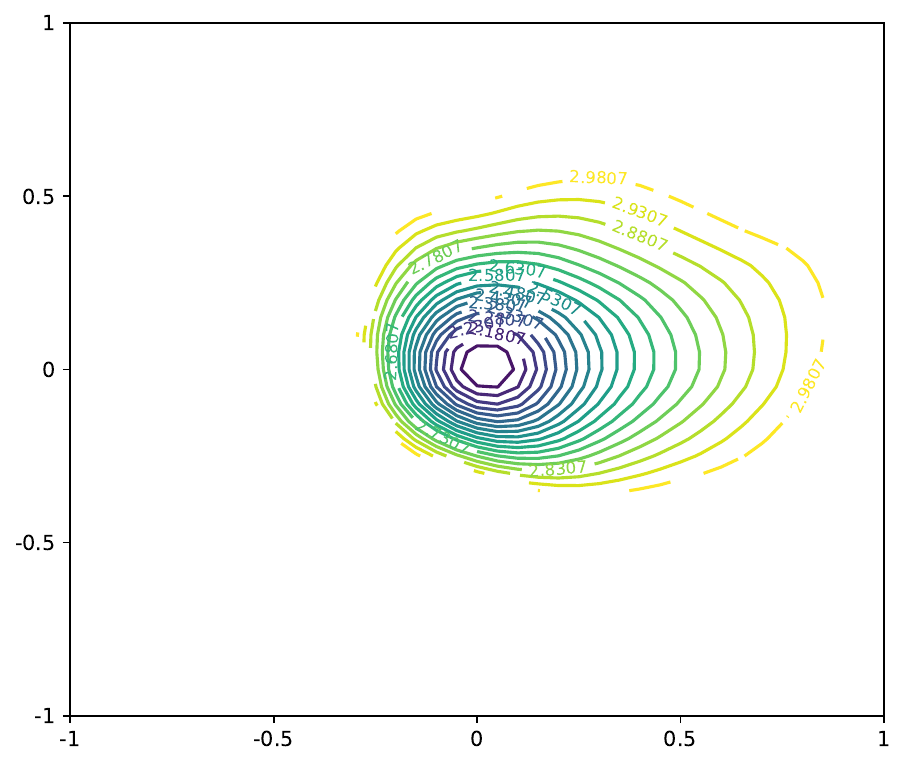}}
\subfigure[Task3, PTUPCDR]{\includegraphics[width=0.21\linewidth]{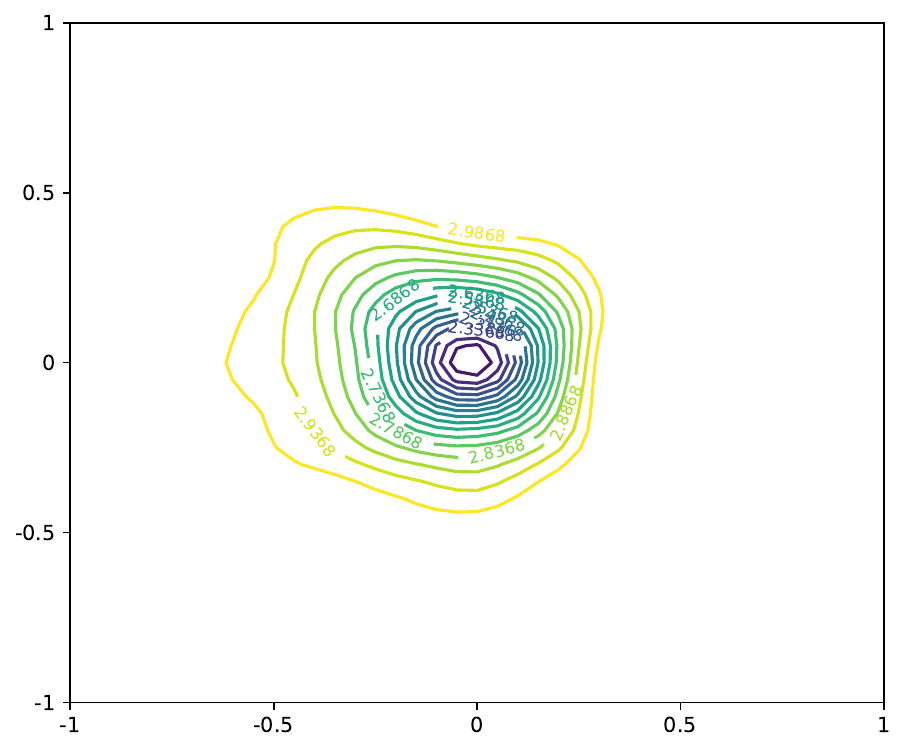}}
\subfigure[Task3, TMCDR]{\includegraphics[width=0.21\linewidth]{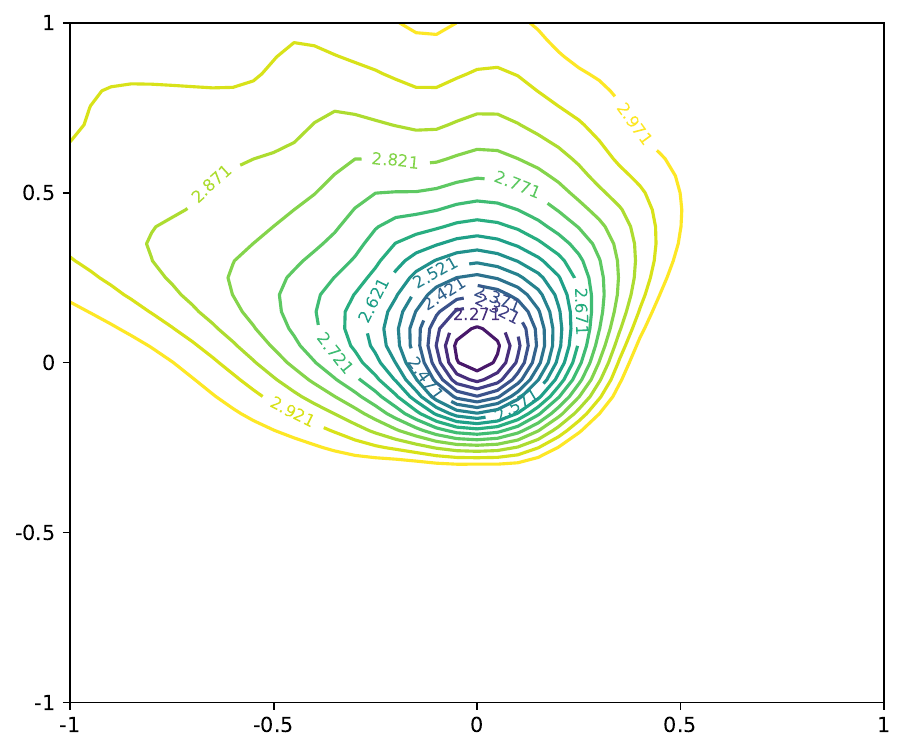}}
\subfigure[Task3, SCDR]{\includegraphics[width=0.21\linewidth]{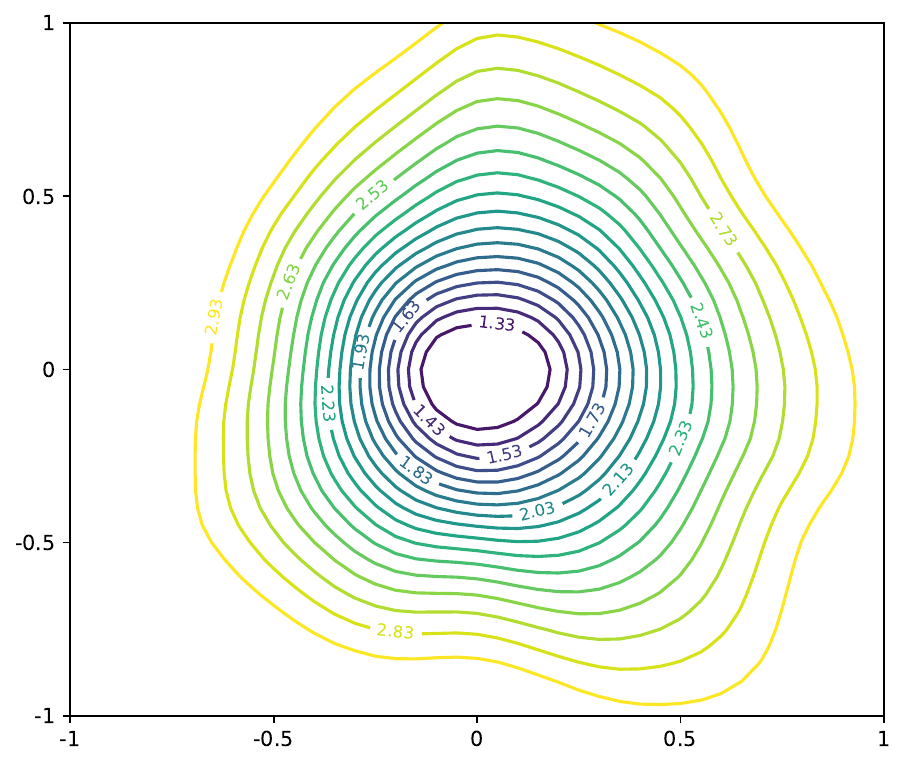}}
\end{tabular}
\caption{Visualization of loss landscape for representation space around $\hat{\mathbf{u}}^s$. $O$ denotes the ratio of overlapping users.}

\label{fig:vis_landscape}
%\vspace{-5mm}
\end{figure*}

\subsection{Loss Landscape Properties of SCDR (RQ2)}

Loss landscape characterizes how a model's loss changes with respect to its parameters \cite{hochreiter1997flat, li2018visualizing}, which is useful for understanding optimization difficulties such as local minima and saddle points. In this section, we aim to visualize the loss landscape of CDR models to observe and analyze their generalization ability. Considering that the EMCDR-method utilizes the representation of overlapping users in the source domain to solve the cold-start problem, we visualize the loss landscape corresponding to the source representation space where overlapping users are located. Note that user representations reside in a high-dimensional vector space, the loss landscape cannot be directly observed. We adopt the visualization method in \cite{li2018visualizing} to obtain a two-dimensional projection of the high-dimensional loss landscape. Specifically, we generate a series of loss values by interpolating along random directions in the trained weight space and map these loss values to a two-dimensional plane, forming a visualized loss landscape.  Algorithm \ref{alg-vis} demonstrates the specific details of the visualization method.

\begin{algorithm}[]    
    \caption{Visualization of CDR 2D loss landscape}
    \begin{algorithmic}[1]
        \Require{sample size $N$, scalar parameters $\zeta \in\left[\zeta_{\min }, \zeta_{\max }\right]$, $\gamma \in\left[\gamma_{\min }, \gamma_{\max }\right]$, The mapping function $f_U$, testing samples $\{\hat{{\mathbf{u}}}^s_i, \hat{\mathbf{v}}^t_j, R_{ij}^t\}^N$ }
        \Ensure{CDR 2D Loss Landscape for $\hat{\mathbf{u}}^s$}
        \State Sample first random direction $\mathbf{g}_1 \sim \mathcal{N}(0,1)$
        \State Sample second random direction $\mathbf{g}_2 \sim \mathcal{N}(0,1)$
        \For{$\zeta$ = $\zeta_{min}$ , \ldots,  $\zeta_{max}$}
        \For{$\gamma$ = $\gamma_{min}$ , \ldots,  $\gamma_{max}$}
            \For{i = 0 ,\ldots , N}
                \State $\delta_1 \leftarrow \frac{\mathbf{g}_{1}}{\left\|\mathbf{g}_{1}\right\|_F}\left\|\hat{\mathbf{u}}_i^s\right\|_F $  
                \State $\delta_2 \leftarrow \frac{\mathbf{g}_{2}}{\left\|\mathbf{g}_{2}\right\|_F}\left\|\hat{\mathbf{u}}_i^s\right\|_F$ 
                \State $\hat{\mathbf{u}}^t_i \leftarrow f_U(\hat{\mathbf{u}}^t_i + \gamma \delta_1+ \zeta \delta_2)$
                \State $y \leftarrow \Braket{\hat{\mathbf{u}}^t_i,\hat{\mathbf{v}}^t_j}$
            \EndFor
            \State $\rho(\hat{\mathbf{u}}^s+\zeta \mathbf{g}_1+ \gamma\mathbf{g}_2) = \frac{1}{N} \sum^N \lvert R_{ij} - y\rvert$
        \EndFor
        \EndFor
        \State Plot 2D Loss Landscape $(\zeta$, $\gamma$, $ \rho(\hat{\mathbf{u}}^s+\zeta \mathbf{g}_1+ \gamma \mathbf{g}_2)), \forall \zeta \in\left[\zeta_{\min }, \zeta_{\max }\right], \forall \gamma \in\left[\gamma_{\min }, \gamma_{\max }\right]$
    \end{algorithmic}
    \label{alg-vis}
\end{algorithm}

Figure \ref{fig:vis_landscape} shows the loss landscapes of the EMCDR, PTUPCDR, TMCDR and the SCDR method using contour plots. The denser the contour lines, the sharper the loss landscape. Based on Figure, we can draw the following conclusions: (1) Compared to the loss landscape of EMCDR, the loss landscape of SCDR is significantly smoother. This indicates that our method can make the model converge to flat minima, suggesting it has better generalization ability \cite{foret2020sharpness}; (2) Additionally, we can observe that there is a certain correlation between the ratio of overlapping users and the steepness of the loss landscape. From Scenario 1 to Scenario 3, as the proportion of overlapping users decreases, the loss landscape becomes increasingly steep. The EMCDR method is more likely to converge to sharp minima in datasets with fewer overlapping users, while our proposed method can avoid converging to sharp minima.

\subsection{Adversarial Robustness Analysis (RQ3)}

Recommender systems have been shown to be vulnerable to adversarial attacks that lead to the model making incorrect recommendations  \cite{anelli2021adversarial, deldjoo2021survey} . In CDR tasks, adversarial samples are assumed to be maliciously perturbed user-item rating pairs, and a robust CDR model should still be able to make accurate predictions for these adversarial samples. Recent research has found a non-trivial connection between SAM and adversarial robustness. \cite{wei2023sharpness} show that SAM can improve a model's adversarial robustness without sacrificing accuracy compared to standard training, which indicates that SCDR may be a robust CDR method.

In this section, we conduct an empirical study to explore how our method exhibits better robustness compared to existing CDR methods. We employed the Fast Gradient Sign Method (FGSM) \cite{DBLP:journals/corr/GoodfellowSS14} to conduct white-box adversarial attacks on the CDR model. FGSM is an adversarial attack technique that generates adversarial examples by adding perturbations to input data in the direction of the gradient of the loss function, exploiting the model's sensitivity to small input changes. 

As a gradient-based attack method, FGSM constructs an adversarial example $\boldsymbol{u}^\prime$ by $\boldsymbol{u}^\prime = \boldsymbol{u} + \boldsymbol{\eta}$, where $\boldsymbol{u}$ denotes the original user representation and $\boldsymbol{\eta}$ is the FGSM perturbation defined by
\begin{equation}
\boldsymbol{\eta}=\epsilon \operatorname{sign}\left(\nabla_{\boldsymbol{u}} \mathcal{L}_{\rm MF}(\mathbf{u},\mathbf{v}) \right) 
\end{equation}
where $\epsilon$ denotes the attack rate of FGSM, $\theta$ is the parameters of a model, $sign()$ denotes the sign function and $\boldsymbol{u}$ is the input to the model. Intuitively, FGSM method perturbs an input by adding noise in the direction of the gradient of the loss with respect to the input, thereby causing the model to misclassify the input data. As the attack rate $\epsilon$ increases, the perturbations to the model become greater, and consequently, the model's performance declines. Table 4 demonstrates the adversarial robustness of SCDR at various attack rates (under Scenario 1, $\beta=80\%$).

\begin{figure}[!h]
\centering
\centerline{\includegraphics[width=0.35\textwidth]{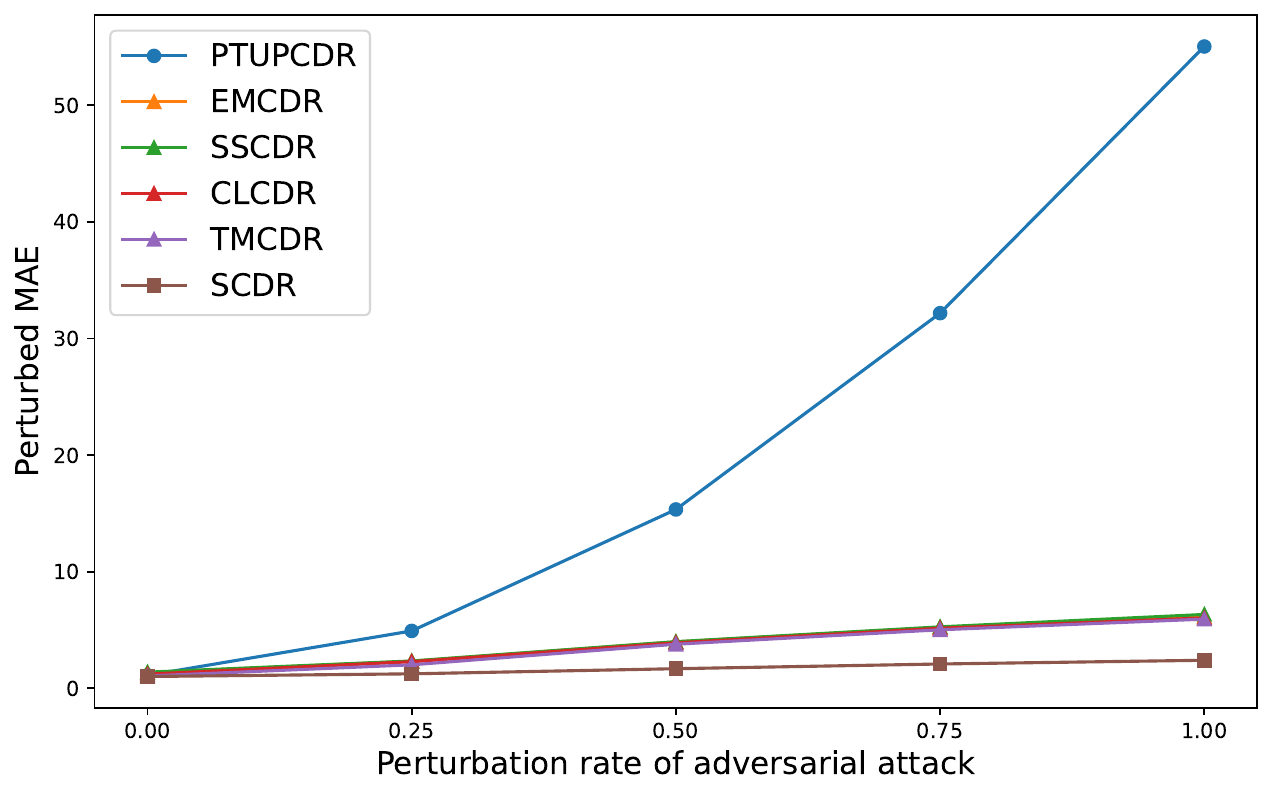}}
%\caption{Example of a figure caption.}
\caption{Adversarial Robustness of CDR methods.}
\label{fig-adv}
\end{figure}

\begin{table}[t]
\centering
\caption{Adversarial Robustness of SCDR Under Different FGSM Attack Rate}
\begin{tabular}{@{}c|c|cccc@{}}
\toprule
\multirow{2}{*}{\textbf{Method}} & \textbf{Natural} & \multicolumn{4}{c}{\textbf{$\ell_2$ Roubst MAE}} \\
                                 & \textbf{MAE}     & $\epsilon$=0.25    & $\epsilon$=0.5    & $\epsilon$=0.75   & $\epsilon$=1      \\ \midrule
SCDR (k=1, $\rho$=1)                  & 1.0246           & 1.4903    & 2.3265   & 3.0282   & 3.5069   \\
SCDR (k=2, $\rho$=1)                  & 1.0218           & 1.4840     & 2.3033   & 2.9669   & 3.4279   \\
SCDR (k=3, $\rho$=1)                  & 1.0208           & 1.4874    & 2.3050    & 2.9765   & 3.4315   \\
SCDR (k=5, $\rho$=1)                  & 1.0167           & 1.4789    & 2.2969   & 2.9576   & 3.4120    \\ \midrule
SCDR (k=1, $\rho$=5)                  & 1.0141           & 1.4888    & 2.3183   & 3.0130    & 3.5174   \\
SCDR (k=2, $\rho$=5)                  & 1.0208           & 1.4241    & 2.2288   & 2.9371   & 3.4648   \\
SCDR (k=3, $\rho$=5)                  & 1.0157           & 1.3362    & 2.0135   & 2.6259   & 3.1078   \\
SCDR (k=5, $\rho$=5)                  & 1.0127          & 1.2413    & 1.6808   & 2.0845   & 2.4044   \\ \bottomrule
\end{tabular}
\end{table}

Figure \ref{fig-adv} shows that all CDR methods exhibit a certain degree of vulnerability to adversarial attacks. As the attack rate increases, the model’s accuracy decreases accordingly. Among them, SDCR demonstrates the best robustness, allowing it to maintain a relatively good recommendation accuracy when subjected to adversarial attacks.  Interestingly, we found that although PTUPCDR has impressive performance, it is very vulnerable to adversarial attacks. This might be due to the  personalized mapping function using more parameters to  learn each user’s personalized mapping, with this over-parameterised neural network structure leading to an unexpectedly fragile nature. In contrast, SCDR robustifies the model against adversarial attacks by smoothing the loss landscape.

\subsection{Hyperparameters Analysis (RQ4)}

In SCDR, there are two important hyperparameters $\rho$ and $k$, where $\rho$ is the maximum radius of the perturbation in Eq. (\ref{eq-inner}) and Eq. (\ref{eq-smf}), and $k$ is the number of iterations in Eq. (\ref{eq-adv1}). Intuitively, a larger $\rho$ represents the strength of sharpness regularization, while a larger $k$ yields a more accurate estimation of sharpness. In this section, we analyze the impact of hyperparameters $\rho$ and $k$ on the model’s performance, robustness, and loss landscape. Before starting the hyperparameter analysis, we introduce a quantitative analysis of the loss landscape rather than merely visualizing it. Specifically, we consider using the Lipschitz constant to measure the landscape.

\subsubsection{Lipschitz constant evaluation}

We use the Lipschitz constant \cite{terjek2019adversarial} to quantitatively describe the sharpness of the loss landscape. Specifically, consider a function $f: X \rightarrow Y$, its Lipschitz constant is defined as
\begin{equation}
\|f\|_L=\sup _{x, y \in X ; x \neq y} \frac{d_Y(f(x), f(y))}{d_X(x, y)} 
\label{eq-lipschitz}
\end{equation}
Eq. (\ref{eq-lipschitz}) characterizes the sharpness of the loss landscape since it quantifies the maximum rate at which the function's values can change, effectively measuring how steep or abrupt changes are within the landscape. Intuitively, a lower Lipschitz constant implies a smoother loss landscape. 

However, accurately calculating the Lipschitz constant over the entire input domain is computationally expensive, since it requires intensive sampling of the function's loss values. Inspired by \cite{terjek2019adversarial}, we leverage a carefully chosen sampling strategy for $x$ and $y$ that can enhance computational efficiency while providing a more accurate estimate of the Lipschitz constant. We rewrite Eq. (17) with $y = x + r$ as
\begin{equation}
\|f\|_L=\sup _{x, x+r \in X ; 0<d_X(x, x+r)} \frac{d_Y(f(x), f(x+r))}{d_X(x, x+r)}
\label{eq-lipschitz2}
\end{equation}
Eq. (\ref{eq-lipschitz2}) assumes that the supremum can be achieved for some perturbation $r$. Given the context of cross-domain recommendation, we let $x$ to be the representation for overlapping users $\hat{\mathbf{u}}^s$, perturbation $r$ to be the $\mathbf{\delta}^*$,
then the Lipschitz constant can be estimated by
\begin{equation}
\|f\|_L^{\prime} = \sup _{\|\boldsymbol{\delta}\|_2 \leq \rho} \mathbb{E}_{\hat{\mathbf{u}}^s \sim P} \frac{|f_U(\hat{\mathbf{u}}^s)-f_U(\hat{\mathbf{u}}^s+\boldsymbol{\delta})|}{\|\hat{\mathbf{u}}^s-(\hat{\mathbf{u}}^s+\boldsymbol{\delta})\|_2} 
\label{eq-lipschitz3}
\end{equation}
where $f_U(\hat{\mathbf{u}}^s)$ denotes the final output of $\hat{\mathbf{u}}^s$ and $P$ denotes the distribution of $\hat{\mathbf{u}}^s$. Specifically, Eq. (\ref{eq-lipschitz3}) explicitly calculates the pairwise Lipschitz constant of the mapping function for the overlapping user $\hat{\mathbf{u}}^s$ and its neighbors within the $\ell_2$ norm ball. Intuitively, Eq. (\ref{eq-lipschitz3}) represents the maximum slope of the loss landscape centered at $\hat{\mathbf{u}}^s$  with a radius of $\rho$. Note that Eq. (\ref{eq-lipschitz3}) directly uses the adversarial samples obtained from Eq. (12) to estimate the Lipschitz constant with pre-calculated $\mathbf{\delta}^*$, thus alleviating the heavy computational cost when explicitly estimate Lipschitz constant with Eq. (\ref{eq-lipschitz}).

\subsubsection{Sensitivity analysis} Now we analyze the impact of hyperparameters on the model’s performance, robustness, and loss landscape. The experiments is conducted in Scenario 1 with $\beta=80\%$. We analyze the performance of SCDR when $\rho$=1 and $\rho$=5, and when $k$ takes values from the set \{1,2,3,5\}.

\begin{figure}[!h]
\centering
\subfigure[SCDR ($\rho=1$)]{
        \includegraphics[width=0.43\columnwidth]{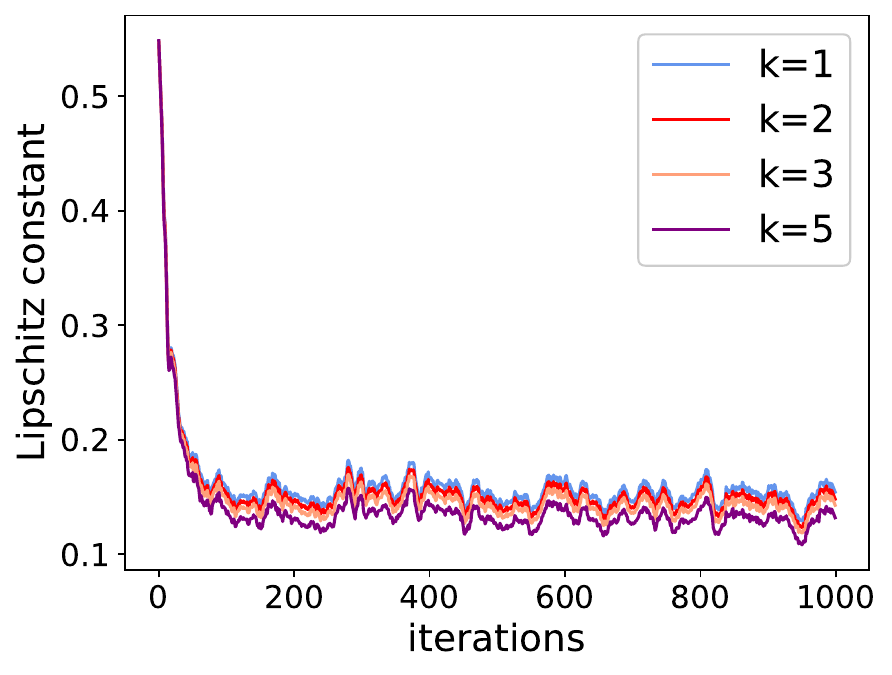}
    }
\subfigure[SCDR ($\rho=5$)]{
        \includegraphics[width=0.43\columnwidth]{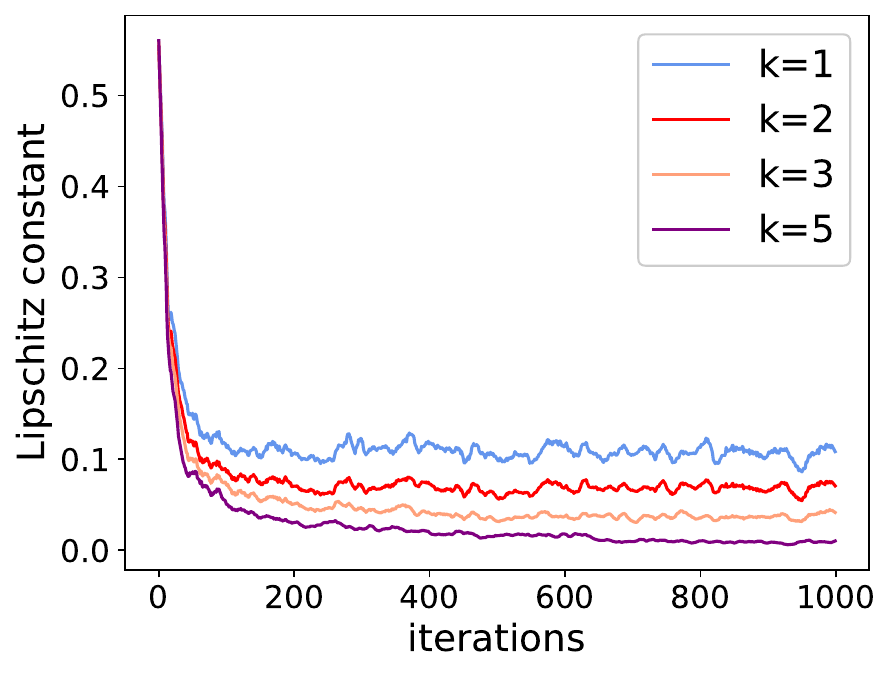}
    }

\caption{The effect of $\rho$ and $k$ on loss landscape sharpness}
\label{fig:effect_sharpness}
\end{figure}

\begin{figure}[!h]
\vspace{-1.4em}
\centering
\subfigure[SCDR ($\rho=1$)]{
        \includegraphics[width=0.43\columnwidth]{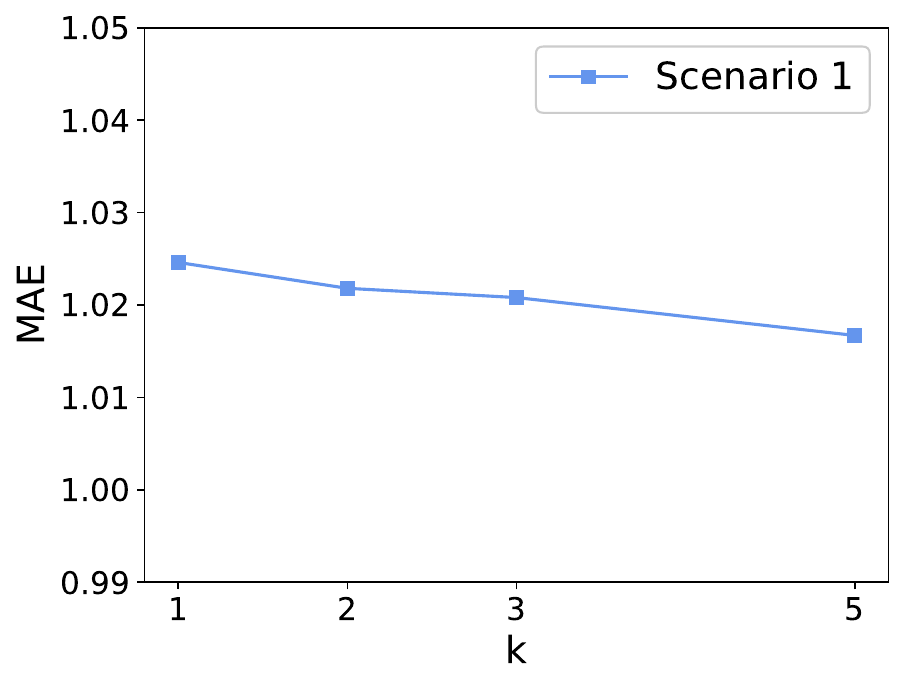}
    }
\subfigure[SCDR ($\rho=5$)]{
        \includegraphics[width=0.43\columnwidth]{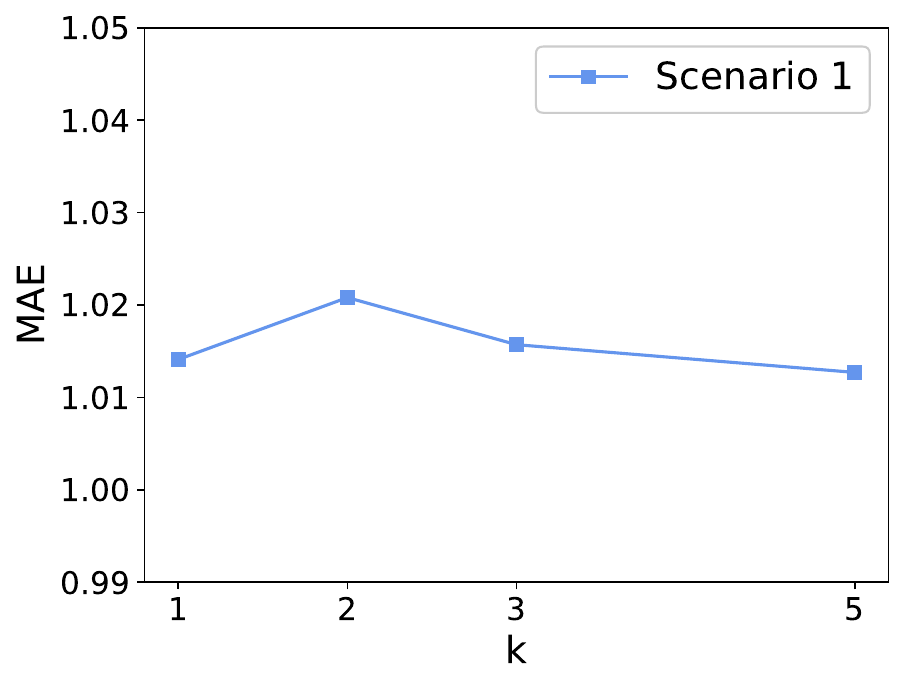}
    }

\caption{The effect of $\rho$ and $k$ on performance}
\label{fig:effect_performance}
\end{figure}

We first analyzed whether larger values of $\rho$ and $k$ could lead to a better approximation of $\delta$ that maximizes $\mathcal{L}_{SCDR}$, thereby more effectively altering the geometric properties of the loss landscape. Figure \ref{fig:effect_sharpness} shows that as $\rho$ and $k$ increase, SCDR is able to converge to flatter local minima. This experimental result aligns with our intuition. Furthermore, as shown in Figure \ref{fig:effect_performance}, we observed that the model's performance is not sensitive to changes in $\rho$ and $k$ which is consistent with SAM's \cite{foret2020sharpness} experimental results on CIFAR-10.

\begin{figure}[!h]
\vspace{-1.4em}
\centering
\subfigure[SCDR ($\rho=1$)]{
        \includegraphics[width=0.43\columnwidth]{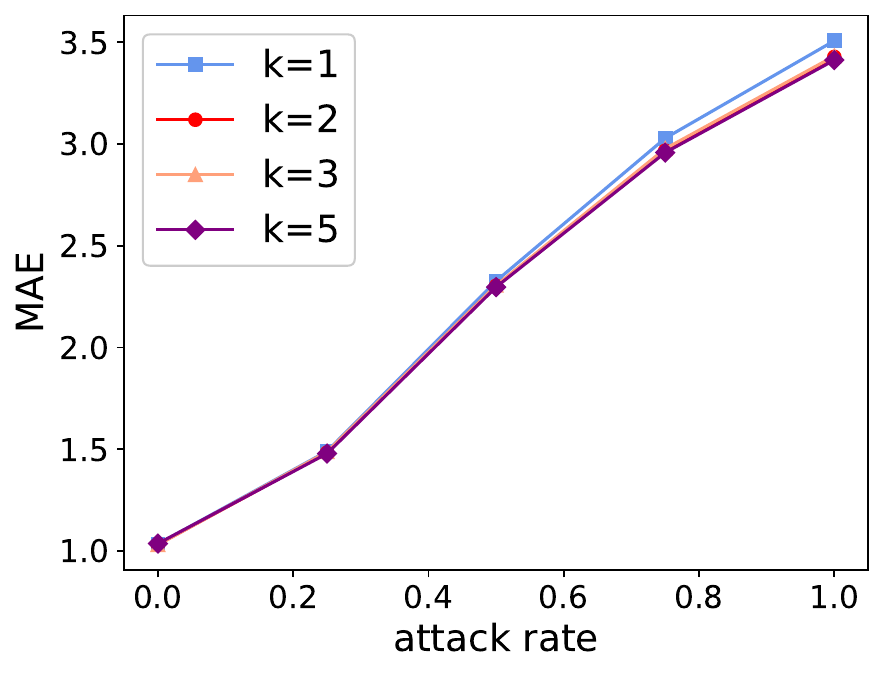}
    }
\subfigure[SCDR ($\rho=5$)]{
        \includegraphics[width=0.43\columnwidth]{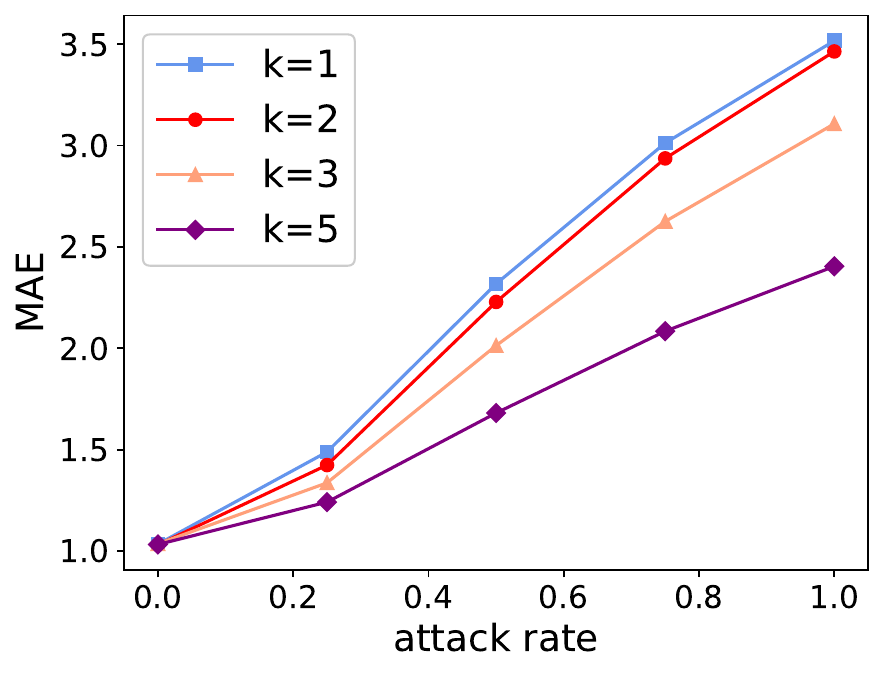}
    }
\caption{The effect of $\rho$ and $k$ on adversarial robustness}
\label{fig:effect_robustness}
\end{figure}

Although the model's performance is not significantly sensitive to $\rho$ and $k$, we found that they have a significant effect on the robustness of SCDR. As shown in Figure \ref{fig:effect_robustness}, when $\rho$ is sufficiently large, a larger $k$ can make the loss landscape flatter and improve the adversarial robustness of the model. This may be because when the model's loss landscape is flatter, it is more difficult to attack the model by gradient-based adversarial attacks. Based on these observations, we can draw the following conclusion: (1) The hyperparameters $\rho$ and $k$ directly affect the sharpness of the loss landscape. Larger values of $\rho$ and $k$ can make the model converge to flatter minima, which is directly related to the model's adversarial robustness. When the model converges to flatter minima, it has stronger robustness against gradient-based adversarial attacks. (2) The recommendation performance of SCDR is not strongly affected by the choice of hyperparameters $\rho$ and $k$.

\subsection{SCDR with other recommendation models (RQ5)}
In SCDR, we employ the matrix factorization (MF) as the backbone recommendation model for generating user and item embeddings. We choose MF as the backbone because other EMCDR-based methods also utilize MF as their recommendation model; this choice helps us avoid introducing additional variables to ensure fair experimental results. In this section, we demonstrate that our method can be applied to the popular graph-based models, such as NGCF \cite{wang2019neural} and LightGCN \cite{he2020lightgcn}.

\begin{table}[!h]
\centering
\caption{SCDR with various recommendation models}
\begin{tabular}{@{}lcccccc@{}}
\toprule
\multirow{2}{*}{\textbf{Method}} & \multicolumn{2}{c}{$\beta$=20\%} & \multicolumn{2}{c}{$\beta$=50\%} & \multicolumn{2}{c}{$\beta$=80\%} \\ \cmidrule(l){2-7} 
                        & \textit{MAE}          & \textit{RMSE}         & \textit{MAE}          & \textit{RMSE}         & \textit{MAE}          & \textit{RMSE}         \\ \midrule
MF (SCDR)                  & 0.9658                & 1.2649                & 0.9729                & 1.2731                & 1.0127                & 1.3316                 \\
NGCF                     & 0.9632                & 1.2631               & 0.9686                & 1.2719                & 1.0262                & 1.3215                \\ 
LightGCN                    &  \textbf{0.9520}               & \textbf{1.2546}              & \textbf{0.9590}                & \textbf{1.2656}                & \textbf{1.0233}                & \textbf{1.3141}                \\ \bottomrule
\end{tabular}
\label{table-various}

\end{table}

Table \ref{table-various} shows the performance when we replace matrix factorization with NGCF and LightGCN as the recommendation models. The experiments are conducted on CDR Scenario 1. Intuitively, the motivation behind SCDR is independent of the choice of recommendation models, and thus it is not limited by the selection of recommendation models. Experimental results show that our proposed SCDR method can be applied to other recommendation models.

\subsection{Ablation Study (RQ6)}

In this section, we conduct an ablation study to validate the effectiveness of SCDR. The experiments are conducted on CDR Scenario 1. First, we set k=0 for the SMF module in the pretraining phase (meaning that SAM is not effective at this point), and we call the resulting model $SCDR^{-}$. Furthermore, we set k=0 for the SCDR module in the learning mapping function part, noting that the model is equivalent to TMCDR \cite{zhu2021transfer} at this point. From the results in Table \ref{table-ablation}, we can draw the following conclusions: (1) Models containing the SAM mod consistently outperform those without the SAM module, indicating the effectiveness of the proposed methods; and (2) In the most extreme sparsity case, SCDR brings the most significant improvement. This suggests that our method can effectively overcome the challenge of overlapping user scarcity.

\begin{table}[!h]
\centering
\caption{Ablation Study}
\begin{tabular}{@{}lcccccc@{}}
\toprule
\multirow{2}{*}{\textbf{Method}} & \multicolumn{2}{c}{$\beta$=20\%} & \multicolumn{2}{c}{$\beta$=50\%} & \multicolumn{2}{c}{$\beta$=80\%} \\ \cmidrule(l){2-7} 
                        & \textit{MAE}          & \textit{RMSE}         & \textit{MAE}          & \textit{RMSE}         & \textit{MAE}          & \textit{RMSE}         \\ \midrule
TMCDR                    & 1.0536                & 1.4164                & 1.2026                & 1.6512                & 1.5872                & 2.1870                \\
$SCDR^{-}$                    & 1.0468                & 1.4046               & 1.1734                & 1.6209                & 1.5266                & 2.1097                \\ \midrule
$SCDR$                    & 0.9658                & 1.2649               & 0.9729                & 1.2731                & 1.0127                & 1.3316                \\ \bottomrule
\end{tabular}
\label{table-ablation}

\end{table}

\section{Conclusion}

In this paper, we propose Sharpness-Aware Cross-Domain Recommendation, namely SCDR, to address the cold-start problem in recommender systems. We observe that existing CDR methods are likely to converge to sharp local minima because only a very limited number of overlapping users are available for training. Based on this observation and inspired by recent studies in sharpness-aware minimization, we propose SCDR to overcome the aforementioned challenge and enhance model generalization. Experiments on the Amazon CDR datasets have validated the superior performance of SCDR and demonstrated that SCDR has better robustness to defend against adversarial attacks.

In future work, we will consider exploring other applications of cross-domain recommendation and the geometric properties of loss, such as in the healthcare domain and applications involving genetic information. Also, when calculating SAM, we neglected the Hessian matrix of the loss function to reduce the computational burden of SCDR. In the future, we will explore how to improve the efficiency of SCDR. Further, we investigate how sharpness-aware minimization can enhance robustness against white-box adversarial attacks. In the future, we plan to extend our research to the robustness against black-box adversarial attacks, where no knowledge of the target model is available.

%\section{Acknowledgements}
%The work presented in this paper was supported by the Australian Research Council (ARC) under Laureate project FL190100149 and discovery project DP220102635.

%\bibliographystyle{plain}
%\bibliography{reference}

\begin{thebibliography}{10}

\bibitem{almazro2010survey}
Dhoha Almazro, Ghadeer Shahatah, Lamia Albdulkarim, Mona Kherees, Romy Martinez, and William Nzoukou.
\newblock A survey paper on recommender systems.
\newblock {\em arXiv preprint arXiv:1006.5278}, 2010.

\bibitem{anelli2021adversarial}
Vito~Walter Anelli, Yashar Deldjoo, Tommaso DiNoia, and Felice~Antonio Merra.
\newblock Adversarial recommender systems: Attack, defense, and advances.
\newblock In {\em Recommender systems handbook}, pages 335--379. Springer, 2021.

\bibitem{bobadilla2013recommender}
Jes{\'u}s Bobadilla, Fernando Ortega, Antonio Hernando, and Abraham Guti{\'e}rrez.
\newblock Recommender systems survey.
\newblock {\em Knowledge-based systems}, 46:109--132, 2013.

\bibitem{chen2023sharpness}
Huiyuan Chen, Chin-Chia~Michael Yeh, Yujie Fan, Yan Zheng, Junpeng Wang, Vivian Lai, Mahashweta Das, and Hao Yang.
\newblock Sharpness-aware graph collaborative filtering.
\newblock In {\em Proceedings of the 46th International ACM SIGIR Conference on Research and Development in Information Retrieval}, pages 2369--2373, 2023.

\bibitem{chen2022clcdr}
Yanyu Chen, Yao Yao, and Wai Kin~Victor Chan.
\newblock Clcdr: Contrastive learning for cross-domain recommendation to cold-start users.
\newblock In {\em International Conference on Neural Information Processing}, pages 331--342. Springer, 2022.

\bibitem{cremonesi2011cross}
Paolo Cremonesi, Antonio Tripodi, and Roberto Turrin.
\newblock Cross-domain recommender systems.
\newblock In {\em 2011 IEEE 11th International Conference on Data Mining Workshops}, pages 496--503. Ieee, 2011.

\bibitem{deldjoo2021survey}
Yashar Deldjoo, Tommaso~Di Noia, and Felice~Antonio Merra.
\newblock A survey on adversarial recommender systems: from attack/defense strategies to generative adversarial networks.
\newblock {\em ACM Computing Surveys (CSUR)}, 54(2):1--38, 2021.

\bibitem{dinh2017sharp}
Laurent Dinh, Razvan Pascanu, Samy Bengio, and Yoshua Bengio.
\newblock Sharp minima can generalize for deep nets.
\newblock In {\em International Conference on Machine Learning}, pages 1019--1028. PMLR, 2017.

\bibitem{du2022sharpness}
Jiawei Du, Daquan Zhou, Jiashi Feng, Vincent Tan, and Joey~Tianyi Zhou.
\newblock Sharpness-aware training for free.
\newblock {\em Advances in Neural Information Processing Systems}, 35:23439--23451, 2022.

\bibitem{dziugaite2017computing}
Gintare~Karolina Dziugaite and Daniel~M. Roy.
\newblock Computing nonvacuous generalization bounds for deep (stochastic) neural networks with many more parameters than training data.
\newblock In Gal Elidan, Kristian Kersting, and Alexander Ihler, editors, {\em Proceedings of the Thirty-Third Conference on Uncertainty in Artificial Intelligence, {UAI} 2017, Sydney, Australia, August 11-15, 2017}. {AUAI} Press, 2017.

\bibitem{foret2020sharpness}
Pierre Foret, Ariel Kleiner, Hossein Mobahi, and Behnam Neyshabur.
\newblock Sharpness-aware minimization for efficiently improving generalization.
\newblock In {\em International Conference on Learning Representations}, 2020.

\bibitem{DBLP:journals/corr/GoodfellowSS14}
Ian~J. Goodfellow, Jonathon Shlens, and Christian Szegedy.
\newblock Explaining and harnessing adversarial examples.
\newblock In Yoshua Bengio and Yann LeCun, editors, {\em 3rd International Conference on Learning Representations, {ICLR} 2015, San Diego, CA, USA, May 7-9, 2015, Conference Track Proceedings}, 2015.

\bibitem{gope2017survey}
Jyotirmoy Gope and Sanjay~Kumar Jain.
\newblock A survey on solving cold start problem in recommender systems.
\newblock In {\em 2017 International Conference on Computing, Communication and Automation (ICCCA)}, pages 133--138. IEEE, 2017.

\bibitem{he2016ups}
Ruining He and Julian McAuley.
\newblock Ups and downs: Modeling the visual evolution of fashion trends with one-class collaborative filtering.
\newblock In {\em proceedings of the 25th international conference on world wide web}, pages 507--517, 2016.

\bibitem{he2020lightgcn}
Xiangnan He, Kuan Deng, Xiang Wang, Yan Li, Yongdong Zhang, and Meng Wang.
\newblock Lightgcn: Simplifying and powering graph convolution network for recommendation.
\newblock In {\em Proceedings of the 43rd International ACM SIGIR conference on research and development in Information Retrieval}, pages 639--648, 2020.

\bibitem{hochreiter1997flat}
Sepp Hochreiter and J{\"u}rgen Schmidhuber.
\newblock Flat minima.
\newblock {\em Neural computation}, 9(1):1--42, 1997.

\bibitem{hu2018conet}
Guangneng Hu, Yu~Zhang, and Qiang Yang.
\newblock Conet: Collaborative cross networks for cross-domain recommendation.
\newblock In {\em Proceedings of the 27th ACM international conference on information and knowledge management}, pages 667--676, 2018.

\bibitem{huang2023boosting}
Bo~Huang, Mingyang Chen, Yi~Wang, Junda Lu, Minhao Cheng, and Wei Wang.
\newblock Boosting accuracy and robustness of student models via adaptive adversarial distillation.
\newblock In {\em Proceedings of the IEEE/CVF Conference on Computer Vision and Pattern Recognition}, pages 24668--24677, 2023.

\bibitem{jiang2019fantastic}
Yiding Jiang, Behnam Neyshabur, Hossein Mobahi, Dilip Krishnan, and Samy Bengio.
\newblock Fantastic generalization measures and where to find them.
\newblock In {\em International Conference on Learning Representations}, 2019.

\bibitem{kang2019semi}
SeongKu Kang, Junyoung Hwang, Dongha Lee, and Hwanjo Yu.
\newblock Semi-supervised learning for cross-domain recommendation to cold-start users.
\newblock In {\em Proceedings of the 28th ACM International Conference on Information and Knowledge Management}, pages 1563--1572, 2019.

\bibitem{kawaguchi2016deep}
Kenji Kawaguchi.
\newblock Deep learning without poor local minima.
\newblock {\em Advances in neural information processing systems}, 29, 2016.

\bibitem{keskar2016large}
Nitish~Shirish Keskar, Dheevatsa Mudigere, Jorge Nocedal, Mikhail Smelyanskiy, and Ping Tak~Peter Tang.
\newblock On large-batch training for deep learning: Generalization gap and sharp minima.
\newblock In {\em 5th International Conference on Learning Representations, {ICLR} 2017}, 2017.

\bibitem{koren2021advances}
Yehuda Koren, Steffen Rendle, and Robert Bell.
\newblock Advances in collaborative filtering.
\newblock {\em Recommender systems handbook}, pages 91--142, 2021.

\bibitem{lai2023enhancing}
Vivian Lai, Huiyuan Chen, Chin-Chia~Michael Yeh, Minghua Xu, Yiwei Cai, and Hao Yang.
\newblock Enhancing transformers without self-supervised learning: A loss landscape perspective in sequential recommendation.
\newblock In {\em Proceedings of the 17th ACM Conference on Recommender Systems}, pages 791--797, 2023.

\bibitem{lam2008addressing}
Xuan~Nhat Lam, Thuc Vu, Trong~Duc Le, and Anh~Duc Duong.
\newblock Addressing cold-start problem in recommendation systems.
\newblock In {\em Proceedings of the 2nd international conference on Ubiquitous information management and communication}, pages 208--211, 2008.

\bibitem{li2018visualizing}
Hao Li, Zheng Xu, Gavin Taylor, Christoph Studer, and Tom Goldstein.
\newblock Visualizing the loss landscape of neural nets.
\newblock {\em Advances in neural information processing systems}, 31, 2018.

\bibitem{lika2014facing}
Blerina Lika, Kostas Kolomvatsos, and Stathes Hadjiefthymiades.
\newblock Facing the cold start problem in recommender systems.
\newblock {\em Expert systems with applications}, 41(4):2065--2073, 2014.

\bibitem{liu2021leveraging}
Weiming Liu, Jiajie Su, Chaochao Chen, and Xiaolin Zheng.
\newblock Leveraging distribution alignment via stein path for cross-domain cold-start recommendation.
\newblock {\em Advances in Neural Information Processing Systems}, 34:19223--19234, 2021.

\bibitem{liu2022towards}
Yong Liu, Siqi Mai, Xiangning Chen, Cho-Jui Hsieh, and Yang You.
\newblock Towards efficient and scalable sharpness-aware minimization.
\newblock In {\em Proceedings of the IEEE/CVF Conference on Computer Vision and Pattern Recognition}, pages 12360--12370, 2022.

\bibitem{lu2015recommender}
Jie Lu, Dianshuang Wu, Mingsong Mao, Wei Wang, and Guangquan Zhang.
\newblock Recommender system application developments: a survey.
\newblock {\em Decision support systems}, 74:12--32, 2015.

\bibitem{madry2017towards}
Aleksander Madry, Aleksandar Makelov, Ludwig Schmidt, Dimitris Tsipras, and Adrian Vladu.
\newblock Towards deep learning models resistant to adversarial attacks.
\newblock In {\em International Conference on Learning Representations}, 2018.

\bibitem{man2017cross}
Tong Man, Huawei Shen, Xiaolong Jin, and Xueqi Cheng.
\newblock Cross-domain recommendation: An embedding and mapping approach.
\newblock In {\em IJCAI}, volume~17, pages 2464--2470, 2017.

\bibitem{mnih2007probabilistic}
Andriy Mnih and Russ~R Salakhutdinov.
\newblock Probabilistic matrix factorization.
\newblock {\em Advances in neural information processing systems}, 20, 2007.

\bibitem{poirier2010reducing}
Damien Poirier, Fran{\c{c}}oise Fessant, and Isabelle Tellier.
\newblock Reducing the cold-start problem in content recommendation through opinion classification.
\newblock In {\em 2010 IEEE/WIC/ACM International Conference on Web Intelligence and Intelligent Agent Technology}, volume~1, pages 204--207. IEEE, 2010.

\bibitem{rafailidis2015modeling}
Dimitrios Rafailidis and Alexandros Nanopoulos.
\newblock Modeling users preference dynamics and side information in recommender systems.
\newblock {\em IEEE Transactions on Systems, Man, and Cybernetics: Systems}, 46(6):782--792, 2015.

\bibitem{singh2008relational}
Ajit~P Singh and Geoffrey~J Gordon.
\newblock Relational learning via collective matrix factorization.
\newblock In {\em Proceedings of the 14th ACM SIGKDD international conference on Knowledge discovery and data mining}, pages 650--658, 2008.

\bibitem{sinha2017certifying}
Aman Sinha, Hongseok Namkoong, and John Duchi.
\newblock Certifying some distributional robustness with principled adversarial training.
\newblock In {\em International Conference on Learning Representations}, 2018.

\bibitem{son2016dealing}
Le~Hoang Son.
\newblock Dealing with the new user cold-start problem in recommender systems: A comparative review.
\newblock {\em Information Systems}, 58:87--104, 2016.

\bibitem{sun2023remit}
Caiqi Sun, Jiewei Gu, BinBin Hu, Xin Dong, Hai Li, Lei Cheng, and Linjian Mo.
\newblock Remit: reinforced multi-interest transfer for cross-domain recommendation.
\newblock In {\em Proceedings of the AAAI Conference on Artificial Intelligence}, volume~37, pages 9900--9908, 2023.

\bibitem{sun2011content}
Dongting Sun, Cong Li, and Zhigang Luo.
\newblock A content-enhanced approach for cold-start problem in collaborative filtering.
\newblock In {\em 2011 2nd international conference on artificial intelligence, management science and electronic commerce (AIMSEC)}, pages 4501--4504. IEEE, 2011.

\bibitem{terjek2019adversarial}
D{\'{a}}vid Terj{\'{e}}k.
\newblock Adversarial lipschitz regularization.
\newblock In {\em 8th International Conference on Learning Representations, {ICLR} 2020}, 2020.

\bibitem{wang2021cross}
Chang-Dong Wang, Yan-Hui Chen, Wu-Dong Xi, Ling Huang, and Guangqiang Xie.
\newblock Cross-domain explicit--implicit-mixed collaborative filtering neural network.
\newblock {\em IEEE Transactions on Systems, Man, and Cybernetics: Systems}, 52(11):6983--6997, 2021.

\bibitem{wang2019neural}
Xiang Wang, Xiangnan He, Meng Wang, Fuli Feng, and Tat-Seng Chua.
\newblock Neural graph collaborative filtering.
\newblock In {\em Proceedings of the 42nd international ACM SIGIR conference on Research and development in Information Retrieval}, pages 165--174, 2019.

\bibitem{wei2023sharpness}
Zeming Wei, Jingyu Zhu, and Yihao Zhang.
\newblock Sharpness-aware minimization alone can improve adversarial robustness.
\newblock In {\em The Second Workshop on New Frontiers in Adversarial Machine Learning}, 2023.

\bibitem{wen2022sharpness}
Kaiyue Wen, Tengyu Ma, and Zhiyuan Li.
\newblock How sharpness-aware minimization minimizes sharpness?
\newblock In {\em The Eleventh International Conference on Learning Representations}, 2022.

\bibitem{wu2019deep}
Di~Wu, Xin Luo, Mingsheng Shang, Yi~He, Guoyin Wang, and MengChu Zhou.
\newblock A deep latent factor model for high-dimensional and sparse matrices in recommender systems.
\newblock {\em IEEE Transactions on Systems, Man, and Cybernetics: Systems}, 51(7):4285--4296, 2019.

\bibitem{wu2020adversarial}
Dongxian Wu, Shu-Tao Xia, and Yisen Wang.
\newblock Adversarial weight perturbation helps robust generalization.
\newblock {\em Advances in Neural Information Processing Systems}, 33:2958--2969, 2020.

\bibitem{xie2022contrastive}
Ruobing Xie, Qi~Liu, Liangdong Wang, Shukai Liu, Bo~Zhang, and Leyu Lin.
\newblock Contrastive cross-domain recommendation in matching.
\newblock In {\em Proceedings of the 28th ACM SIGKDD Conference on Knowledge Discovery and Data Mining}, pages 4226--4236, 2022.

\bibitem{zhang2021deep}
Qian Zhang, Wenhui Liao, Guangquan Zhang, Bo~Yuan, and Jie Lu.
\newblock A deep dual adversarial network for cross-domain recommendation.
\newblock {\em IEEE Transactions on Knowledge and Data Engineering}, 2021.

\bibitem{zhu2019dtcdr}
Feng Zhu, Chaochao Chen, Yan Wang, Guanfeng Liu, and Xiaolin Zheng.
\newblock Dtcdr: A framework for dual-target cross-domain recommendation.
\newblock In {\em Proceedings of the 28th ACM International Conference on Information and Knowledge Management}, pages 1533--1542, 2019.

\bibitem{zhu2021graphical}
Feng Zhu, Yan Wang, Chaochao Chen, Guanfeng Liu, and Xiaolin Zheng.
\newblock A graphical and attentional framework for dual-target cross-domain recommendation.
\newblock In {\em Proceedings of the Twenty-Ninth International Conference on International Joint Conferences on Artificial Intelligence}, pages 3001--3008, 2021.

\bibitem{zhu2021cross}
Feng Zhu, Yan Wang, Chaochao Chen, Jun Zhou, Longfei Li, and Guanfeng Liu.
\newblock Cross-domain recommendation: challenges, progress, and prospects.
\newblock In {\em 30th International Joint Conference on Artificial Intelligence, IJCAI 2021}, pages 4721--4728. International Joint Conferences on Artificial Intelligence, 2021.

\bibitem{zhu2017chrs}
Junxing Zhu, Jiawei Zhang, Chenwei Zhang, Quanyuan Wu, Yan Jia, Bin Zhou, and S~Yu Philip.
\newblock Chrs: Cold start recommendation across multiple heterogeneous information networks.
\newblock {\em IEEE Access}, 5:15283--15299, 2017.

\bibitem{zhu2021transfer}
Yongchun Zhu, Kaikai Ge, Fuzhen Zhuang, Ruobing Xie, Dongbo Xi, Xu~Zhang, Leyu Lin, and Qing He.
\newblock Transfer-meta framework for cross-domain recommendation to cold-start users.
\newblock In {\em Proceedings of the 44th international ACM SIGIR conference on research and development in information retrieval}, pages 1813--1817, 2021.

\bibitem{zhu2022personalized}
Yongchun Zhu, Zhenwei Tang, Yudan Liu, Fuzhen Zhuang, Ruobing Xie, Xu~Zhang, Leyu Lin, and Qing He.
\newblock Personalized transfer of user preferences for cross-domain recommendation.
\newblock In {\em Proceedings of the Fifteenth ACM International Conference on Web Search and Data Mining}, pages 1507--1515, 2022.

\end{thebibliography}

\vfill

\end{document}